\documentclass[11pt]{article}
\pdfoutput=1 
\usepackage{jheppub} 
\usepackage[utf8]{inputenc}
\usepackage{braket,slashed,bm}
\usepackage{array,multirow}
\usepackage{amsmath}
\usepackage[normalem]{ulem}
\usepackage[T1]{fontenc}
\usepackage{mathrsfs}
\usepackage{tikz}
\newcommand{\abs}[1]{\left|#1\right|}
\usepackage{subcaption}
\usetikzlibrary{arrows,decorations.pathmorphing,backgrounds,positioning,fit,petri,automata,shadows,calendar,mindmap,
decorations.markings,calc}
\usepackage{lscape}
\usepackage{hyperref}
\usepackage{graphicx}
\usepackage{booktabs}
\usepackage{floatrow}
\newfloatcommand{capbtabbox}{table}[][\FBwidth]
\def\beq{\begin{equation}}
\def\eeq{\end{equation}}
\def\bea{\begin{eqnarray}}
\def\eea{\end{eqnarray}}

\def\hyp{\mathsf{y}}
\renewcommand{\a}{\alpha}

\renewcommand{\d}{\delta}
\newcommand{\g}{\gamma}
\newcommand{\s}{\sigma}
\newcommand{\de}{\partial}
\newcommand{\LL}{\mathcal{L}}

\newcommand{\Q}{\mathcal{Q}}
\newcommand{\CHlt}{C_{Hl}^{(3)}}
\newcommand{\CHqt}{C_{Hq}^{(3)}}
\newcommand{\CHls}{C_{Hl}^{(1)}}
\newcommand{\CHqs}{C_{Hq}^{(1)}}

\newcommand{\st}{s_\theta}

\graphicspath{{../}}

\title{Scheming in the SMEFT\dots\\
and a reparameterization invariance!}
\author{
Ilaria Brivio and Michael Trott\\
Niels Bohr International Academy and Discovery Center, Niels Bohr Institute,
University of Copenhagen, Blegdamsvej 17, DK-2100 Copenhagen, Denmark}
\abstract{We explain a reparameterization invariance in the Standard Model Effective Field Theory present 
when considering $\bar{\psi} \psi \rightarrow \bar{\psi} \psi$ scatterings (with $\psi$ a fermion) and how this leads to unconstrained combinations of Wilson coefficients in global data analyses restricted to these measurements.
We develop a $\{\hat{m}_W, \hat{m}_Z,\hat{G}_F\}$ input parameter scheme and compare results to the case when an input parameter set $\{\hat{\alpha}, \hat{m}_Z,\hat{G}_F\}$
is used to constrain this effective theory from the global data set, confirming the input parameter independence of the
unconstrained combinations of Wilson coefficients, and supporting the reparameterization  
invariance explanation. We discuss some conceptual issues related to these degeneracies that are relevant for LHC data reporting and analysis.}
\begin{document}
\maketitle

\section{Introduction} \label{sec:intro}

Hidden invariances can be present in Effective Field Theories (EFTs) and
explain empirically observed structures of the EFT, or relations between otherwise free parameters in the theory.
These relations and structures are important to uncover when the Standard Model (SM) is promoted to the Standard Model Effective Field Theory (SMEFT) in order to
systematically search for the effects of physics beyond the SM. When such physics
is present in corrections to SM predictions, significant phenomenological consequences can result.\footnote{Examples of non-intuitive aspects of SMEFT phenomenology based on the subtle structure of this field theory include a unitarity and helicity based understanding of the one loop approximate Holomorphy in the $\mathcal{L}_6$ Renormalization Group  \cite{Alonso:2014rga,Cheung:2015aba}, non-interferences in tree level scatterings due to helicity \cite{Azatov:2016sqh,Falkowski:2016cxu}, 
and the global symmetry based structure embedded in the
SMEFT operator expansion \cite{deGouvea:2014lva}.}

An empirically observed structure of the SMEFT
is how the constraints from a large set of pre-LHC data project onto the Wilson coefficients of higher dimensional operators. Two unconstrained directions in the SMEFT Wilson coefficient space have been found in the global $\bar{\psi} \psi \rightarrow \bar{\psi} \psi$ data set. This fact is manifest in the particular operator basis of Ref.~\cite{Grzadkowski:2010es}, but not in other formalisms.
The incorporation of $\bar{\psi} \psi \rightarrow \bar{\psi} \psi \, \bar{\psi} \psi$ scattering data is known to lift these unconstrained directions \cite{Hagiwara:1993ck,DeRujula:1991ufe}, so it is critical to incorporate this data in order to globally constrain the SMEFT parameters leading to anomalous $Z$ couplings.

In this paper we explain how the presence of unconstrained directions in $\bar\psi\psi\to\bar\psi\psi$ scattering data is due to the fact that the description of 
these processes is invariant under a particular reparameterization, which is illustrated in detail in Section~\ref{sec:reparam}. In Section \ref{sec:reparamscaling} we discuss how $\bar\psi\psi\to\bar\psi\psi\,\bar\psi\psi$ scattering data breaks this structure because it does not respect the same invariance in an operator basis independent manner using a scaling argument. 

The reparameterization invariance is not due to a symmetry of the SMEFT, but rather originates as a property of $\bar\psi\psi\to\bar\psi\psi$ scattering processes. As such, it is always present as a basis independent feature of this class of measurements. Nonetheless, how this translates into the appearance of unconstrained directions in a global fit analysis does depend on the operator basis employed. We discuss the issue of UV assumptions and basis choice, and how utilizing a mass eigenstate formalism or various power counting assumptions can make the impact of the reparameterization invariance non-manifest in Section \ref{sec:basisissues}.

The interpretation of this invariance is subtle because it requires the equations of motion (EOM) to understand and, further, it is a property limited to a subset of observables used to define the numerical values
of the Lagrangian parameters through "input parameters". As a consequence, one could speculate that these unconstrained directions are just accidental structures related to a particular basis or input parameter set. 
In order to examine the input parameter scheme dependence, we perform a global data analysis for LEPI data on the properties of the $Z$ boson, $e^+ e^- \rightarrow e^+ e^-$ scattering and  $e^+ e^-  \rightarrow \bar{\psi} \psi  \bar{\psi} \psi$ production data in the $\{\hat{\alpha}, \hat{m}_Z,\hat{G}_F\}$ and in the $\{\hat{m}_W, \hat{m}_Z,\hat{G}_F\}$
input parameter schemes. In Section \ref{fitresults} we demonstrate that these results confirm the input parameter independence of the reparameterization invariance.
At the same time, the correlations and constraints on operators due to observables of different Feynman diagram topologies, 
even in the same operator basis, show some numerical scheme dependence, as we also show. These results also support
assigning a SMEFT theoretical error to naive leading order global constraint analyses, as we discuss in Section  \ref{fitresults}.

Finally, in Section \ref{conclusions}, we conclude with some comments on the impact of these results on SMEFT analyses of global
data sets including LHC data.
\subsection{The Standard Model Effective Field Theory} \label{sec:SMEFTintro}
The EFT approach to physics beyond the SM introduces local contact operators to capture the low energy, or infrared (IR), limit of such
physics below new physics scale(s) $\sim \Lambda$.  When the following assumptions are also made, this approach has come to be known as the SMEFT.

First, it is assumed that  $\rm SU(2)_L \times U(1)_Y$ is spontaneously broken to $\rm U(1)_{em}$ by the vacuum expectation value ($\langle H^\dagger H \rangle  \equiv \bar{v}_T^2/2$) of the Higgs doublet field. Second, the observed scalar is assumed to be $J^P = 0^+$ and embedded in a doublet of $\rm SU(2)_L$ in the EFT construction. Thereby, 
no large non-linearities are introduced by ultraviolet (UV) dynamics that is integrated out, which distinguishes this
approach from the HEFT (Higgs-EFT) formalism \cite{Feruglio:1992wf,Burgess:1999ha,Grinstein:2007iv,Buchalla:2012qq,Alonso:2012px,Buchalla:2013rka,Brivio:2013pma,Brivio:2014pfa,Brivio:2016fzo,Alonso:2014wta,Alonso:2015fsp,Alonso:2016oah}.
Finally, the SMEFT also assumes a mass gap so that $\bar{v}_T/\Lambda<1$. The $\mathcal{L}_{SMEFT}$ that follows is the sum of $\rm SU(3)_C \times SU(2)_L \times U(1)_Y$ invariant higher dimensional operators built out of SM fields
\bea
\mathcal{L}_{SMEFT} = \mathcal{L}_{SM} + \mathcal{L}^{(5)} + \mathcal{L}^{(6)} + \mathcal{L}^{(7)} + ..., \quad \quad \mathcal{L}^{(k)}= \sum_{\alpha = 1}^{n_k} \frac{C_\alpha^{(k)}}{\Lambda^{k-4}} Q_\alpha^{(k)} \hspace{0.25cm} \text{ for $k > 4$. }
\eea
Here $\mathcal{L}^{(k)}$ contains the dimension $k$ operators $Q_\alpha^{(k)}$. The number of non redundant operators in $\mathcal{L}^{(5)}$, $\mathcal{L}^{(6)}$, $\mathcal{L}^{(7)}$ and $\mathcal{L}^{(8)}$ is known \cite{Buchmuller:1985jz,Grzadkowski:2010es,Weinberg:1979sa,Abbott:1980zj,Lehman:2015via,Lehman:2014jma,Lehman:2015coa,Henning:2015alf}. We adopt a naive power counting in mass dimension in this paper.
This choice makes the reparameterization invariance clearer as we discuss in Section \ref{sec:PCissues}, where we also comment on the impact of alternative operator normalization choices.
We employ the Warsaw basis of dimension six operators of Ref.~\cite{Grzadkowski:2010es} with the notation $Q_i$ to denote an operator 
defined in this basis. See Ref.~\cite{Grzadkowski:2010es} for the explicit operator definitions. 
We use a notation where we implicitly absorb the factor of $1/\Lambda^2$ into the $C_i$ for most results, unless explicitly noted.
Further notational conventions are the use of a hat superscript for input parameters, or Lagrangian parameters related to input parameters at tree level,
and a bar superscript for canonically normalized $\mathcal{L}_{SMEFT}$ parameters.
\section{Reparameterization invariance in the SMEFT} \label{sec:reparam}
When considering small perturbations to SM predictions in an EFT
it is required to clearly distinguish a signal process used  to uncover such perturbations from background processes. 
Frequently, this signal isolation is done by exploiting tree level resonant exchange with a minimum number of initial states and final states, so that the signal process is kinematically distinct enough from the background in how it populates phase space to be well measured.

This practical experimental consideration makes 
$\bar{\psi} \psi \rightarrow V \rightarrow \bar{\psi} \psi$ scattering a critical process to make precise measurements in a collider environment.
Here $\psi$ are spin $1/2$ states, so that the intermediate state is spin one or spin zero, and we 
consider the case that $V$ is a vector field.
The reparameterization invariance at work in the SMEFT in $\bar{\psi} \psi \rightarrow \bar{\psi} \psi$ scattering\footnote{The same invariance is also present in a subset of the diagrams contributing to other scattering processes. An example related to $\bar\psi\psi\to\bar\psi\psi\,\bar\psi\psi$ scattering is shown in Section~\ref{sec:reparamscaling}. } 
is due to the degeneracy in the normalization of the kinetic term of $V$ and $V \bar{\psi} \psi$ corrections when considering these processes. 
Consider the following schematic Lagrangian of $d \leq 4$ interactions
\bea\label{schematicvector}
\mathcal{L}_{V \psi_i} = \frac{1}{2} \, m_V^2 \, V^\mu \, V_\mu - \frac{1}{4}V^{\mu \, \nu} V_{\mu \, \nu}  - g \, \bar{\psi}_i \gamma^\mu \psi_j V_\mu - g \, \kappa \, \bar{\psi}_k \gamma^\mu \psi_l V_\mu + \cdots.
\eea
where $V^{\mu \, \nu} = \partial^\mu \, V^\nu - \partial^\nu \, V^\mu$ and $m_V\propto g$. Here $i,j,k,l$
are flavour indices.
It is not important that the coupling of the vector field to the fermions $\psi$ is normalized to be the same (as indicated by the rescaling by $\kappa$ in the last term), only that the couplings are both 
proportional to the same parameter.  
The vector boson can always be transformed between canonical and non-canonical form in its kinetic term by a field redefinition without physical effect 
due to a corresponding correction in the LSZ formula \cite{Lehmann:1954rq}.\footnote{A naive treatment of a massive vector boson as an asymptotic  $S$ matrix element can also introduce challenges from gauge invariance, however, see the discussion in Ref.~\cite{Passarino:2010qk}, and references therein, on how this naivety can be avoided.}
Restricting one's attention to the interactions explicit in Eqn.~\ref{schematicvector}, such a shift can be canceled by a corresponding shift of $g$. This fact is used in standard formulations of the SMEFT to take the theory to canonical form, where
correlated transformations of the form $g_b \rightarrow g_b' (1- \epsilon)$ and $V_b \rightarrow V_b' (1+ \epsilon)$, with $\epsilon \sim \bar{v}_T^2/\Lambda^2$, that leave $g_b \, V_b \rightarrow g'_b \, V'_b$ invariant for $V_b = \{G,B,W\}$ are used\footnote{See for example the discussion in Section 5.4 of Ref.~\cite{Alonso:2013hga}.}.
The freedom to make these transformations also defines an unobservable physical redundancy of description in a subset of scattering events. The same set of physical 
 $\bar{\psi} \psi \rightarrow V \rightarrow  \bar{\psi} \psi$ scatterings at tree level can be parameterized by an equivalence class of fields and coupling parameters
 \bea\label{WOW}
 \left(V,g \right) \leftrightarrow \left(V' \, (1+ \epsilon), g' \, (1- \epsilon) \right),
 \eea
where  $\epsilon \sim \mathcal{O}(\bar{v}_T^2/\Lambda^2)$. 
This is clearly reminiscent of reparameterization invariance in Heavy Quark Effective Field Theory \cite{Dugan:1991ak,Luke:1992cs}, and as a result we will refer to this
as SMEFT reparameterization invariance. 

This invariance is present when considering a subclass of observables\footnote{For further discussions on reparameterization invariance in EFT's see Refs~\cite{Manohar:1993qn,Manohar:2002fd,Neubert:1993mb,Heinonen:2012km,Hill:2012rh}. } due to
the condition that the amplitude derived is proportional to the same power of $g$ and $V$ rescalings.
This invariance has a physical impact through the EOM relations between classes of operators that have been discussed in the literature
a number of times before in Refs.~\cite{SanchezColon:1998xg,Kilian:2003xt,Grojean:2006nn,Alonso:2013hga} although 
its understanding in terms of an operator basis independent reparameterization invariance has not been discussed in detail previously. 

In this identification of a reparameterization invariance we have neglected the effect of $m_\psi^2/m_V^2$ corrections (we have used Feynman gauge above)
and numerically suppressed terms and loop corrections.
The degeneracy of description is present so long as these effects are neglected.
For example, for this class of $S$ matrix elements a condition is that $m_\psi^2/m_V^2 \ll C_i \, v^2/\Lambda^2$.
This is a good approximation for $V = \{W,Z\}$  for $\Lambda$
in the few {\rm TeV} range. Neglecting SMEFT loop corrections when considering LEPI near $Z$ pole data is not advisable \cite{Passarino:2012cb,Berthier:2015oma,Berthier:2015gja,Berthier:2016tkq,deFlorian:2016spz,Passarino:2016pzb,Hartmann:2016pil}.
Nevertheless, we demonstrate in this paper how the unconstrained directions present in naive leading order analyses come about due to this invariance.

\subsection{EOM implementation of the reparameterization invariance} \label{sec:EOMreparam}
The consequences of the reparameterization invariance require the use of the EOM to fully explore.
The SM EOM that are relevant are 
\begin{align}
\left[D^\alpha , W_{\alpha \beta} \right]^I &= g_2  j_\beta^I, & \quad
D^\alpha B_{\alpha \beta} &= g_1  j_\beta ,
\label{eomX}
\end{align}
where $\left[D^\alpha , W_{\alpha \beta} \right]$ is the covariant derivative in the adjoint representation for a vector field tensor $W_{\alpha \beta}$.
The $\rm SU(2)_L$ field and coupling are $(W,g_2)$ and the $\rm U(1)_Y$ field and coupling are $(B,g_1)$.
We use $I,J,K$ for $\rm SU(2)_L$ indices and $i,j,k,l...$ for fermion flavour indices.
The electroweak gauge currents are
\begin{equation}
\begin{aligned}
j_\beta^I &= \frac 12 \overline q \, \tau^I \gamma_\beta  q + \frac12 \overline l \, \tau^I \gamma_\beta  l + \frac12 H^\dagger \, i\overleftrightarrow D_\beta^I H\,, \\
j_\beta &= \sum_{\substack{\psi_\kappa=u,d,\\ 
q,e,l}} \hyp_k  \, \overline \psi_\kappa \, \gamma_\beta  \psi_\kappa + \frac12 H^\dagger \, i\overleftrightarrow D_\beta H\,,
\end{aligned}
\end{equation}
where $\hyp_k$ are the $\rm U(1)_Y$ hypercharges of the fermions, $q$ and $l$ are $\rm SU(2)_L$ left handed doublet fermion fields. The Hermitian derivatives are
\begin{equation}
\begin{aligned}
H^\dagger \, i\overleftrightarrow D_\beta H &= i H^\dagger (D_\beta H) - i (D_\beta H)^\dagger H \,,\\
H^\dagger \, i\overleftrightarrow D_\beta^I H &= i H^\dagger \tau^I (D_\beta H) - i (D_\beta H)^\dagger\tau^I H,\,
\end{aligned}
\end{equation} 
with $\tau^I$ the Pauli matrix. From the EOM, the following operator identities can be obtained \cite{Kilian:2003xt,Grojean:2006nn,Grzadkowski:2010es,Alonso:2013hga}
\begin{subequations}\label{EOMrelations_complete}
\begin{align}
\hyp_h \, g_1^2 Q_{HB} &=  g_1^2 \, j_\beta \, (H^\dagger \, i\overleftrightarrow D_\beta H) - \frac{1}{2} g_1 \, g_2 \, Q_{HWB} + 2 \, i \, g_1 (D_\mu H)^\dagger (D_\nu H) B^{\mu \, \nu}, \label{EOMrelations_complete_B}\\
 g_2^2 Q_{HW} &=  2 \, g_2^2 \, j^I_\beta \, (H^\dagger \, i\overleftrightarrow D_\beta^I H) - 2 \, g_1 \, g_2 \, \hyp_h \, Q_{HWB} + 4 \, i \, g_2 (D_\mu H)^\dagger \tau^I (D_\nu H) W_I^{\mu \, \nu}.\label{EOMrelations_complete_W}
\end{align}
\end{subequations}
We now denote by $S_R$ the class of $\bar{\psi} \psi \rightarrow \bar{\psi} \psi $ matrix elements, which are consistent with the reparameterization invariance of Eqn.~\ref{WOW}. When projecting into this specific category of processes, the following relations are obtained:
\begin{subequations}\label{EOMrelations}
\begin{align}
\langle \hyp_h \, g_1^2 Q_{HB} \rangle_{S_R} &= \langle \sum_{\substack{\psi_\kappa=u,d,\\ 
q,e,l}} \hyp_k \, g_1^2 \, \overline \psi_\kappa \, \gamma_\beta  \psi_\kappa \, (H^\dagger \, i\overleftrightarrow D_\beta H) 
+   \frac{g_1^2}{2} \left(Q_{H \Box}  + 4 Q_{HD}\right)
- \frac{1}{2} g_1 \, g_2 \, Q_{HWB}  \rangle_{S_R}, \label{EOMrelations_B}\\
\langle  \, g_2^2 Q_{HW} \rangle_{S_R} &= \langle  g_2^2 \,(\overline q \, \tau^I \gamma_\beta  q + \overline l \, \tau^I \gamma_\beta  l ) \, (H^\dagger \, i\overleftrightarrow D_\beta^I H) 
+  2 \, g_2^2 \, Q_{H \Box} 
- 2 \, g_1 \, g_2 \, \hyp_h \, Q_{HWB}  \rangle_{S_R}. \label{EOMrelations_W}
\end{align}
\end{subequations}
Here $\langle \cdots \rangle_{S_R}$ indicates the projection of operators into the subclass of matrix elements defined above. When applied on Eqs.~\ref{EOMrelations_complete}, the projection selects the operators that do contribute at tree level to the $S_R$ processes and it removes the other ones. In this case, the operators of the form $(D_\mu H)^\dagger X^{\mu\nu} D_\mu H$ (where $X=\{B,W\}$) were removed because they affect triple gauge couplings (TGC) and Higgs-gauge couplings.
The effect of $Q_{H \Box}$ can also be neglected in our case, although formally present through $\bar{\psi} \psi \rightarrow  h \rightarrow \bar{\psi} \psi$, as it is further suppressed by small Yukawa couplings and by the ratio $\Gamma_Z/m_Z$ when considering near $Z$ pole LEPI data.
For the $\bar{\psi} \psi \rightarrow \bar{\psi} \psi$ scatterings of interest we have
\bea\label{kineticterms}
\langle \hyp_h \, g_1^2 Q_{HB} \rangle_{S_R} \rightarrow \frac{g_1^2 \, \bar{v}_T^2}{4 \, \Lambda^2} \, B^{\mu \, \nu} \, B_{\mu \, \nu}, \quad 
\langle g_2^2 Q_{HW} \rangle_{S_R} \rightarrow \frac{g_2^2 \, \bar{v}_T^2}{2 \, \Lambda^2} \, W_I^{\mu \, \nu} \, W^I_{\mu \, \nu}.
\eea
Because of the reparameterization invariance, these structures are not observable in $\bar{\psi} \psi \rightarrow \bar{\psi} \psi$ scatterings. 
The invariance of $S$ matrix elements under field configurations equivalent by use of the EOM implies, then, that this must also hold for the fixed linear combinations of operators appearing on the right-hand sides of Eqs.~\ref{EOMrelations}.
In the Wilson coefficient space, this translates into the fact that $\bar{\psi} \psi \rightarrow \bar{\psi} \psi$ scattering data alone cannot access neither the coefficients $C_{HB}$, $C_{HW}$ nor the two combinations
\begin{subequations}\label{wBW}
\begin{align}
 g_1^2 \, w_B &= g_1^2\frac{\bar{v}_T^2}{\Lambda^2}\left(
 -\frac{1}{3}C_{Hd}- C_{He}-\frac{1}{2} \CHls+\frac{1}{6} \CHqs+\frac{2}{3} C_{Hu}+2C_{HD}-\frac{1}{2t_{\hat\theta}}C_{HWB}
 \right),\\
 g_2^2 \, w_W &= g_2^2\frac{\bar{v}_T^2}{\Lambda^2}\left(
 \frac{\CHqt+ \CHlt}{2}-\frac{t_{\bar\theta}}{2}C_{HWB}
 \right).
\end{align}
\end{subequations}
The $S_R$ class of data is simultaneously invariant under the two independent reparameterizations that leave the products $(g_1 B_\mu)$ and $(g_2W^i_\mu)$ unchanged, so that the vectors $w_B$ and $w_W$ constitute a basis for the vector space of unconstrained directions. This result holds independently of whether the operators $Q_{HB}$ and $Q_{HW}$ themselves are present or not in the chosen $\mathcal{L}_6$ basis.

Using the global fit described in Section \ref{fitresults} under the assumption of zero SMEFT theoretical error \cite{Passarino:2012cb,Berthier:2015oma,Berthier:2015gja,deFlorian:2016spz,Passarino:2016pzb},
the unconstrained directions in the $\{\hat{\alpha}, \hat{m}_Z,\hat{G}_F\}$ scheme are found to be
\begin{subequations}\label{empiracalflatalpha}
\begin{align}
w^{\a}_1&= \frac{\bar{v}_T^2}{\Lambda^2} \left(\frac{1}{3}C_{Hd}-2C_{HD}+ C_{He}+\frac{1}{2} \CHls-\frac{1}{6} \CHqs-\frac{2}{3} C_{Hu}-1.29 (\CHqt+ \CHlt)+ 1.64 C_{HWB}\right),\\
w^{\a}_2&=\frac{\bar{v}_T^2}{\Lambda^2} \left(\frac{1}{3}C_{Hd}-2C_{HD}+ C_{He}+\frac{1}{2} \CHls-\frac{1}{6} \CHqs-\frac{2}{3} C_{Hu}+ 2.16 (\CHqt+ \CHlt)- 0.16 C_{HWB}\right). 
\end{align}
\end{subequations}
These can be projected into the vector space defined by $w_{B,W}$ as
\bea\label{basisdecomp}
w^{\a}_1= -w_B - 2.59 \, w_W & \quad \quad w^{\a}_2= -w_B +4.31 \, w_W.
\eea
This result is consistent with these unconstrained directions having their origin in a reparameterization invariance.

The physical consequences of these unconstrained directions are subtle. A
direct matching onto the SMEFT from a UV sector is unlikely to correspond to exactly the unconstrained directions in 
$\bar{\psi} \psi \rightarrow \bar{\psi} \psi$ data in the following sense. So long as the operators $Q_{HB}$ and $Q_{HW}$ are retained in the basis they are likely to receive such a
direct matching contribution. The unconstrained directions make
manifest the need to measure Feynman diagrams of different topologies than $\bar{\psi} \psi \rightarrow V   \rightarrow\bar{\psi} \psi$
in order to constrain the properties of the gauge bosons vertex corrections to the SM fermions consistently as the Wilson coefficients of individual operators can carry different meanings in different operator bases. As a result, the fit spaces of EFT approaches to physics beyond the SM are
expected to be intrinsically highly correlated across measurement classes. This is exactly found in global fit results. 
A consequence is the correct treatment of correlations (both theoretical and experimental) between measurements
is critical to obtain a consistent global constraint picture. We return to this point below.

\subsection{Basis choices and reparameterization invariance in the SMEFT}\label{sec:basisissues}
When constructing a complete, independent operator basis, Eqs.~\ref{EOMrelations_complete} are employed to remove two among the operators appearing in those expressions from the final chosen set. In particular, Eq.~\ref{EOMrelations_complete_B} allows to remove one among $Q_{HB}$, $D_\mu H^\dagger B^{\mu\nu}D_\nu H$ and the fermionic invariants with a $\rm SU(2)_L$ singlet contraction, while Eq.~\ref{EOMrelations_complete_W} allows to remove one among $Q_{HW}$, $D_\mu H^\dagger W^{\mu\nu}D_\nu H$ and the fermionic invariants with a $\rm SU(2)_L$ triplet contraction. For the sake of illustrating the physical interpretation of the reparameterization invariance, we explore here the consequences of three alternative choices\footnote{For previous discussions see Ref.~\cite{deGouvea:2014lva,Kilian:2003xt,Grojean:2006nn,Grzadkowski:2010es,Alonso:2013hga}.}:

\begin{itemize}
 \item choosing to remove $D_\mu H^\dagger X^{\mu\nu}D_\nu H$, $X=\{B,W\}$ as in the Warsaw basis~\cite{Grzadkowski:2010es}.
 
Since the operators removed do not affect $\bar\psi\psi\to\bar\psi\psi$ scatterings at tree level, the reparameterization invariance belonging to these processes manifests itself as the presence of four unconstrained parameters: $C_{HW}$, $C_{HB}$ and the two combinations $w_B$, $w_W$ defined in~\ref{wBW}. The inclusion of $\bar\psi\psi\to\bar\psi\psi\,\bar\psi\psi$ data lifts the degeneracies within $w_B$, $w_W$ but leaves $C_{HW}$, $C_{HB}$ unconstrained.
 
 \item choosing to remove $Q_{HB}$, $Q_{HW}$.
 
 As above, the analysis of $\bar\psi\psi\to\bar\psi\psi$ scatterings leaves four quantities unconstrained:  $w_B$, $w_W$ and the Wilson coefficients assigned to the two $D_\mu H^\dagger X^{\mu\nu}D_\nu H$ operators. Including $\bar\psi\psi\to\bar\psi\psi\,\bar\psi\psi$ data allows to access two out of these four, but because all the Wilson coefficients considered contribute to the latter processes, the two residual unconstrained directions shall be linear combinations of the initial four.
 
 \item choosing to remove two fermionic invariants while retaining all the bosonic operators, as in the case of the construction reported in Ref.~\cite{Contino:2013kra}.
 
Because the fermionic operators participate in $\bar\psi\psi\to\bar\psi\psi$ processes, in this case the vectors $w_B$, $w_W$ do not have any direct physical meaning. However, there are still four unconstrained parameters in the Z-pole data, namely $C_{HW}$, $C_{HB}$ and the Wilson coefficients of the two $D_\mu H^\dagger X^{\mu\nu}D_\nu H$ operators, and $\bar\psi\psi\to\bar\psi\psi\,\bar\psi\psi$ data allows to access the latter two. In practice, the reparameterization invariance is still present but simply does not manifest itself as the striking presence of two flat directions involving nine Wilson coefficients. In this sense we refer to this scenario as ``hidden invariance''. 
 
\end{itemize}
When considering the last case, it is important to stress that choosing a $\mathcal{L}_6$ operator basis does not automatically give the Wilson coefficients a physical meaning. This occurs when enough measurements are performed and consistently projected onto the field theory so that all Wilson coefficients in a non-redundant basis are experimentally constrained.  In the case of an operator basis choice where the reparameterization invariance is hidden, the operators introduced are naively not of a form that corresponds
to a modification of the vector fermion bilinear couplings, but of a TGC vertex. Any extraction of a TGC vertex experimentally uses asymptotic states where the massive vector bosons have decayed, so this distinction is not relevant for $S$ matrix elements.

Finally, we note the choice of removing fermionic invariants from an operator basis requires some special care due to the presence of flavour indices.
The key point here is that the relations in Eqs.~\ref{EOMrelations} involve complete sums of SM currents,
\bea
\sum_{\substack{\psi_\kappa=u,d,
q,e,l}} \hyp_\kappa \, g_1^2 \, \overline \psi^i_\kappa \, \gamma_\beta  \psi^i_\kappa \, (H^\dagger \, i\overleftrightarrow D_\beta H). 
\eea
The complete sum of currents involves fermion fields that have the flavour index ($i$), on the other hand, the kinetic terms of the 
vector bosons are not flavour dependent. The choice of operator basis does not have a physical effect, so long as 
no flavour symmetry is explicitly broken by an assumption with choosing a basis, and the operator basis used respects the equivalence theorem 
\cite{Kallosh:1972ap,tHooft:1972qbu,Politzer:1980me,Einhorn:2013kja,Passarino:2016pzb,Passarino:2016saj} in its relation to the Warsaw basis,
(i.e. the operator bases should be related by gauge independent field redefinitions). 
Once enough measurements are made and mapped to the SMEFT in a consistent fashion to constrain all parameters,
which requires a combination of Higgs data, $\bar\psi\psi\to\bar\psi\psi$ data and  $\bar\psi\psi\to\bar\psi\psi\,\bar\psi\psi$ data these unconstrained directions
in the Wilson coefficient space can be consistently constrained simultaneously. However, we stress that it is required {\it to not assume correlations or lack thereof between parameters that act to explicitly break the consequences of the reparameterization invariance while doing so to obtain basis independent results.}\footnote{Marginalization over subsets of operator Wilson coefficients with a prior inconsistent with
the physical consequences of the reparameterization invariance is a common way to bias a global analysis.}

\subsubsection{Power counting choices and reparameterization invariance}\label{sec:PCissues}

A number of historical conceptual barriers have blocked this understanding of 
$\bar{\psi} \psi \rightarrow \bar{\psi} \psi$ scatterings in the SMEFT.  Until the discovery of the Higgs like boson,
it was appropriate and well motivated to use the STU approach to electroweak precision data (EWPD) \cite{Kennedy:1988sn,Altarelli:1990zd,Altarelli:1991fk,Golden:1990ig,Holdom:1990tc,Peskin:1990zt,Peskin:1991sw}. This approach was of manifest utility,  but it is not field redefinition invariant and it does not lend itself to this understanding of the reparameterization invariance\footnote{See Ref.~\cite{Grinstein:1991cd,Han:2004az} for initial steps in the operator based EFT approach.}. 

Differing power counting choices can also block this understanding.
In this work we are using a naive power counting in terms of operator mass dimension, which allows the reparameterization invariance to be identified directly. The naive dimensional analysis
power counting scheme discussed in Refs.~\cite{Manohar:1983md,Cohen:1997rt,Luty:1997fk,Gavela:2016bzc} preserves the relations~\ref{EOMrelations}
in the sense that the operators of classes
\bea
X_{\mu \nu} \, X^{\mu \nu} \, H^\dagger H, \quad \quad H^\dagger \overleftrightarrow D^\mu H \, \bar{\psi} \, \gamma_\mu \psi, \quad \quad  D^2\, H^4,
\eea 
are assigned the same power counting. We also note that these operators are assigned the same chiral number, see Ref.~\cite{Gavela:2016bzc}.
Alternative approaches \cite{Buchalla:2013eza,Buchalla:2014eca,Giudice:2007fh,Pomarol:2014dya,Panico:2015jxa} can introduce UV dependence that can prevent the reparameterization invariance from being manifest. 
This is because the relations~\ref{EOMrelations} are a property of the SMEFT when treated as a field theory irrespective of its unknown UV completion,
and UV matchings need not preserve it. 

For example, some operators in the EOM, and in the unconstrained directions $w^{\a}_{1,2}$
have frequently been associated with 
"tree-loop" operator schemes \cite{Arzt:1994gp} and "universal theories" \cite{Barbieri:2004qk,Wells:2015uba,Wells:2015cre}.
The EOM relations in Eqn.~\ref{EOMrelations} directly relate and equate operators
of a tree and loop form in their projection onto $\bar{\psi} \psi \rightarrow \bar{\psi} \psi$ scatterings, so this UV bias is very difficult to reconcile with the reparameterization invariance discussion above.
The idea of universal theories suffers from the same issue, as some operators present
in Eqn.~\ref{EOMrelations} are of a universal form, and others are not.
Despite this, so long as the Wilson coefficients are allowed to counteract such an operator normalization
choice when fitting the data in a global analyses, one can still uncover the unconstrained directions
in the $\mathcal{L}_6$ Wilson coefficient space, no matter what operator normalization is adopted.

A recent approach of using mass eigenstate coupling parameters to characterize deviations from the Standard Model makes the presence of unconstrained directions 
even harder to uncover in data analyses. The EOM relations key to understanding
the reparameterization invariance do not have a (manifest) equivalent in 
 the parameterization chosen, although the fact that there remain un-probed aspects of the $Z$ boson phenomenology in $\bar{\psi} \psi \rightarrow \bar{\psi} \psi$ scatterings is acknowledged in Refs.~\cite{Gupta:2014rxa,Falkowski:2014tna}. It is also worth noting that defining correlations
 for mass eigenstate parameter formalisms in a form that manifestly preserves the consequences of the
 reparameterization invariance (while maintaining a consistent use of the data) in global analyses remains an unsolved problem.

\subsection{Scalings of scatterings to break degeneracies} \label{sec:reparamscaling}
It is  understood that
$\bar{\psi} \psi \rightarrow \bar{\psi} \psi \, \bar{\psi} \psi$ scattering measurements are required to
fully constrain parameters present in LEP data in an unambiguous fashion. This has been observed by examining higher dimensional operator EOM relations,
and also discovered explicitly in global data analyses \cite{Han:2004az}. The
fact that these measurements break the invariance can be understood with the
following simple operator basis independent scaling argument.

Consider scattering  of the form $\bar{\psi} \psi \rightarrow \bar{\psi} \psi \, \bar{\psi} \psi$. The processes shown in $\mathcal{A}_{3}$
are perturbative corrections to the SM interactions in a manner that preserves the reparameterization invariance.
The topology shown in the middle figure, $\mathcal{A}_{2}$ might be considered to be perturbated by the rescaling of the SM kinetic term
of the $\rm SU(2)_L$ field. However, these corrections drop out, in a manner that is consistent with Eqn.\ref{WOW} being preserved, 
which does not lead to a relative shift in the TGC vertex. As a consequence dependence on the operator $Q_{HW}$ cancels in this process.
However, the amplitude $\mathcal{A}_{2}$ can also be perturbed from the SM value by the introduction of the terms 
labeled in the Effective Lagrangian with $g_1^{Z,\gamma}$, $\kappa_{Z,\gamma}$ and $\lambda_{Z,\gamma}$ in Eqn.~\ref{effectivetgc}.
These non-vanishing contributions, not definable as a $W$ or $B$ field rescaling consistent the reparameterization invariance, are not forbidden by any symmetry.
The corresponding amplitudes are directly not invariant under Eqn.\ref{WOW}, due to these unfixed rescaling parameters no matter what basis is chosen.
\begin{figure}\centering
  		\includegraphics[width=0.31\textwidth]{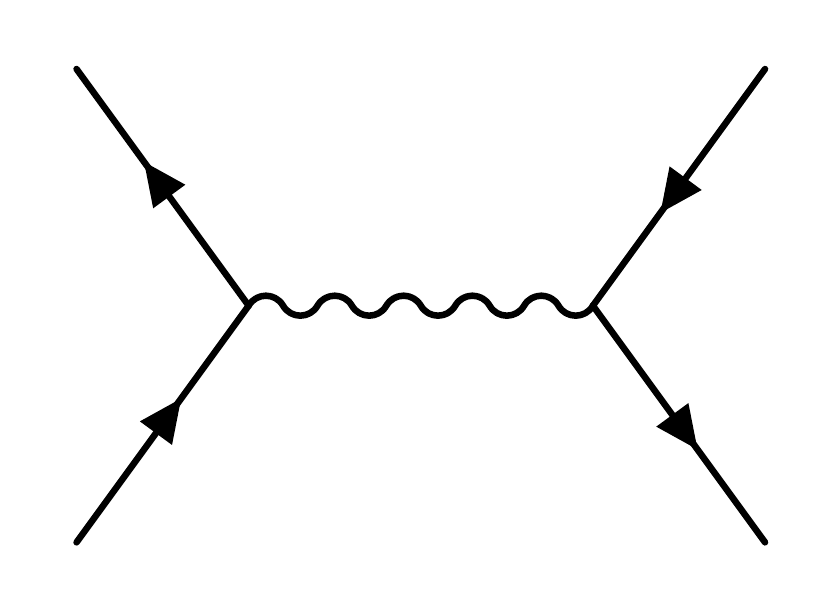}
    		\includegraphics[width=0.31\textwidth]{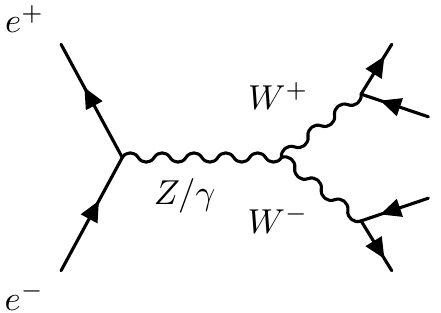}
    		\includegraphics[width=0.31\textwidth]{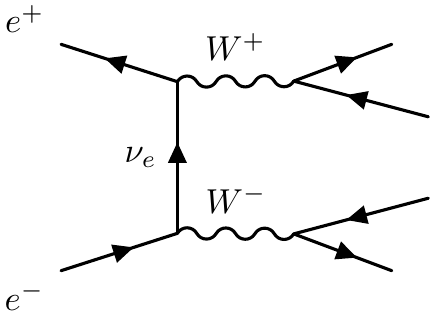}\\[2mm]
	\parbox{\textwidth}{\centering $(\mathcal{A}_1)$\hspace*{.24\textwidth}$(\mathcal{A}_{2Z/2A})$\hspace*{.24\textwidth} ($\mathcal{A}_3$)}
	\caption{Sample diagram topologies for $\bar{\psi} \psi \rightarrow \bar{\psi} \psi$ and doubly resonant $\bar{\psi} \psi \rightarrow \bar{\psi} \psi \, \bar{\psi} \psi$ scattering
	with charged currents. }
	\label{fig:phase_space}
\end{figure}
The degeneracy is weakly broken experimentally as the t-channel neutrino exchange diagram is  
dominant numerically \cite{Hagiwara:1993ck} near the $W^+ \, W^-$ threshold that dominates 
LEPII data.\footnote{For this reason it is also important to break this
degeneracy in a consistent manner by using the $\bar{\psi} \psi \rightarrow \bar{\psi} \psi \, \bar{\psi} \psi$ scattering data, and avoid using 
constraints modeled and projected onto an effective-TGC vertex if possible \cite{Trott:2014dma}. Numerically this issue does not seem
to dramatically effect numerical conclusions comparing the results in \cite{Falkowski:2014tna,Berthier:2016tkq}, although all of these results are
subject to very substantial theoretical uncertainties \cite{Passarino:2012cb,Berthier:2015oma,Berthier:2015gja,Berthier:2016tkq,deFlorian:2016spz,Passarino:2016pzb,Hartmann:2016pil} and the results in \cite{Falkowski:2014tna,Berthier:2016tkq} are so highly constrained they mimic "one at a time" operator analyses
that cannot be consistent with the consequences of the reparameterization invariance.}

\section{Input parameter independence of physical SMEFT conclusions}
A physical reparameterization invariance of $\bar{\psi} \psi \rightarrow \bar{\psi} \psi$ scatterings
should be input parameter set independent.
Input parameters play a critical role
in a perturbative field theory analysis. The parameters present in the Lagrangian (couplings and scales) need to be fixed numerically using a set of precisely measured input observables. Nevertheless, the input parameters are a choice and the existence of a reparameterization invariance and its consequences should not be limited to only one input parameter set.  
The reason is that although the inferred numerically defined Lagrangian used to interpolate between and define $S$ matrix elements in a perturbative expansion introduces an input parameter scheme dependence into predictions, if physical observables are related to each other directly, then this scheme dependence cancels out. In this sense the existence of 
a physical reparameterization invariance should not depend on input parameter choice.

Clearly the individual numerical results present in the global fit do depend upon the input parameter set chosen. 
To complete this argument, it is required to demonstrate the input parameter independence of the reparameterization invariance
in $\bar{\psi} \psi \rightarrow \bar{\psi} \psi$ scatterings by showing that 
a decomposition similar to Eqn.~\ref{basisdecomp} can be performed in the $\{\hat{m}_W, \hat{m}_Z,\hat{G}_F\}$ input scheme.

The $\{\hat{\alpha}, \hat{m}_Z,\hat{G}_F\}$ input scheme is in common use in the literature, so we do not exhaustively discuss this approach here, see Refs.~\cite{Alonso:2013hga,Falkowski:2014tna,Ellis:2014dva,Ellis:2014jta,Berthier:2015oma,Berthier:2015gja,Berthier:2016tkq,Gonzalez-Alonso:2016etj,Cirigliano:2016nyn}. Results directly related to the fit in use here are in Ref.~\cite{Alonso:2013hga,Berthier:2015oma,Berthier:2015gja,Berthier:2016tkq,Hartmann:2016pil} in the SMEFT. We use the numerical values for the input parameters in Table \ref{inputalpha}.
In the next section we develop the $\{\hat{m}_W, \hat{m}_Z,\hat{G}_F\}$ scheme for the SMEFT, while in Appendix~\ref{appendix_heft} we do the same for the HEFT Lagrangian, in the basis of Ref.~\cite{Brivio:2016fzo}.
\begin{center}
\begin{table}
\centering
\tabcolsep 8pt
\begin{tabular}{|c|c|c|}
\hline
Input parameters & Value & Ref. \\ \hline
$\hat{m}_Z$ & $91.1875 \pm 0.0021$ & \cite{ALEPH:2005ab,Agashe:2014kda,Mohr:2012tt} \\
$\hat{G}_F$ & $1.1663787(6) \times 10^{-5} $ &  \cite{Agashe:2014kda,Mohr:2012tt} \\
$\hat{\alpha}_{ew}$ & $1/137.035999074(94) $ &  \cite{Agashe:2014kda,Mohr:2012tt} \\
$\hat{m}_h$ & $125.09 \pm 0.21 \pm 0.11$ & \cite{Aad:2015zhl} \\
$\hat{m}_t$&$173.21 \pm 0.51 \pm 0.71$& \cite{Agashe:2014kda}\\
$\hat{\alpha}_s$&$0.1185$&\cite{Agashe:2014kda}\\
$\Delta \hat{\alpha}$&$0.0590$& \cite{Freitas:2014hra}\\ \hline
\end{tabular}
\caption{Input parameters values used in the global fit in the $\{\hat{\alpha}, \hat{m}_Z,\hat{G}_F\}$ scheme.\label{inputalpha}}
\end{table}
\end{center}
\vspace{-1cm}

\subsection{\texorpdfstring{$\{\hat{m}_W, \hat{m}_Z,\hat{G}_F\}$}{\{mW,mZ,GF\}} input parameter scheme}

{\bf Tree level:} In this scheme, the measured SM Lagrangian parameters are inferred following  the tree level definitions
\begin{align}
\hat{g}_2 &= 2\cdot 2^{1/4}\hat{m}_W\sqrt{\hat{G}_F}, & \quad \hat{g}_1 &= 2\cdot 2^{1/4}\hat{m}_Z\sqrt{\hat{G}_F \left(1 -\frac{\hat{m}_W^2}{\hat{m}_Z^2}\right)}, & \quad \hat{v}^2 &= \frac{1}{\sqrt2 \hat{G}_F},
\end{align}
and in addition
\begin{align}
s^2_{\hat\theta} &= 1-\frac{\hat{m}_W^2}{\hat{m}_Z^2}, & \quad \hat{e} &=  2\cdot 2^{1/4} \hat{m}_W\sqrt{\hat{G}_F} s_{\hat\theta}.
\end{align}

{\bf \noindent Core shifts parameters:} The input parameters are written as their canonically normalized Lagrangian expressions -- $\bar{y}_i$ -- plus a contribution proportional to the relevant $\mathcal{L}_6$ Wilson coefficients, denoted $\d y_i$, so that
\begin{equation}
\hat{y_i} =  \bar{y}_i + \d y_i, \hspace*{1cm} \bar{y}_i=\{\bar{G}_F, \bar{m}_Z^2, \bar{m}_W^2\}\,,
\end{equation} 
and we have in the $\rm U(3)^5$ flavour symmetric limit the results for the input parameter shifts  \footnote{Here we have normalized the operators in $\mathcal{L}_6$ in a manner that does not introduce
a gauge coupling $g_i$ for each field strength tensor. This is the same normalization as used in Refs.~\cite{Alonso:2013hga,Berthier:2015oma,Berthier:2015gja,Berthier:2016tkq}.}

\bea
\d G_F &=& 
\dfrac{1}{\sqrt{2} \, \hat{G}_F}\left(\sqrt2 C_{H\ell}^{(3)}-\dfrac{1}{\sqrt2}C_{ll}\right), \\
\dfrac{\d m_Z^2}{\hat{m}_Z^2} &=& 
\dfrac{1}{2\sqrt2 \hat{G}_F}C_{HD}+\dfrac{\sqrt{2}}{\hat{G}_F}\dfrac{\hat{m}_W}{\hat{m}_Z}\sqrt{1-\dfrac{\hat{m}_W^2}{\hat{m}_Z^2}}C_{HWB}, \\
\dfrac{\d m_W^2}{\hat{m}_W^2} &=& 0.
\eea
In addition we define the short hand notation for the shift in the Weinberg angle in terms of input parameters
\bea
\d s^2_{\theta} &=&\frac{1}{2\sqrt2 \hat{G}_F}\frac{\hat{m}_W^2}{\hat{m}_Z^2}C_{HD}+\frac{1}{\sqrt2 \hat{G}_F}\frac{\hat{m}_W}{\hat{m}_Z}\sqrt{1-\frac{\hat{m}_W^2}{\hat{m}_Z^2}} C_{HWB}.
\eea
{\bf \noindent Effective $Z$ couplings:} The effective axial and vector couplings of the SMEFT $Z$ boson are defined with the normalization
\bea
\mathcal{L}_{Z,eff}  =  2 \, 2^{1/4} \, \sqrt{\hat{G}_F} \, \hat{M}_Z   \left(J_\mu^{Z \ell} Z^\mu + J_\mu^{Z \nu} Z^\mu + J_\mu^{Z u} Z^\mu +  J_\mu^{Z d} Z^\mu \right),
\eea 
where $(J_\mu^{Z x})^{pr} = \bar{x}_p \, \gamma_\mu \left[(\bar{g}^{x}_V)_{eff}^{pr}- (\bar{g}^{x}_A)_{eff}^{pr} \, \gamma_5 \right] x_r$ for $x = \{u,d,\ell,\nu \}$.
Restricting our attention to a minimal $\rm U(3)^5$ linear MFV scenario $(J_\mu^{Z x})_{pr} \simeq (J_\mu^{Z x}) \delta_{pr}$
we define the shifted effective axial and vector couplings as
\bea
\delta (g^{x}_{V,A})_{pr} = (\bar{g}^{x}_{V,A})^{eff}_{pr} - (g^{x}_{V,A})^{SM}_{pr}, 
\eea
and
\begin{align}\label{Zcouplngs.shift}
 \d g_V^f &= \delta \bar{g}_Z \, \bar{g}_{V}^x+ Q^f\d \st^2 + \Delta^f_V,  & \quad  \d g_A^f &= \delta \bar{g}_Z \, \bar{g}_A + \Delta_A^f.
\end{align}
Our normalization of the couplings in $\mathcal{L}_{Z,eff}$ is such that $\bar{g}_{V}^x = T_3/2 - Q^x \, \bar{s}_\theta^2, \bar{g}_A = T_3/2$ where $T_3 = 1/2$ for $u_i,\nu_i$ and $T_3 = -1/2$ for $d_i,\ell_i$
and $Q^x = \{-1,2/3,-1/3 \}$ for $x = \{\ell,u,d\}$. $\Delta^f_{V,A}$ stands for the direct contributions from fermionic operators given by
\begin{align}
 \Delta^\ell_V &= -\frac{1}{4\sqrt 2 \hat{G}_F}\left(C_{H\ell}^{(1)}+C_{H\ell}^{(3)}+C_{He}\right)&
 \Delta^\ell_A &= -\frac{1}{4\sqrt 2 \hat{G}_F}\left(C_{H\ell}^{(1)}+C_{H\ell}^{(3)}-C_{He}\right),\\
 \Delta^\nu_V &= -\frac{1}{4\sqrt 2 \hat{G}_F}\left(C_{H\ell}^{(1)}-C_{H\ell}^{(3)}\right)&
 \Delta^\nu_A &= -\frac{1}{4\sqrt 2 \hat{G}_F}\left(C_{H\ell}^{(1)}-C_{H\ell}^{(3)}\right),\\
 \Delta^u_V &= -\frac{1}{4\sqrt2 \hat{G}_F}\left(C_{Hq}^{(1)}-C_{Hq}^{(3)}+C_{Hu}\right)&
 \Delta^u_A &= -\frac{1}{4\sqrt2 \hat{G}_F}\left(C_{Hq}^{(1)}-C_{Hq}^{(3)}-C_{Hu}\right),\\
 \Delta^d_V &= -\frac{1}{4\sqrt2 \hat{G}_F}\left(C_{Hq}^{(1)}+C_{Hq}^{(3)}+C_{Hd}\right)&
 \Delta^d_A &= -\frac{1}{4\sqrt2 \hat{G}_F}\left(C_{Hq}^{(1)}+C_{Hq}^{(3)}-C_{Hd}\right),
\end{align}
where
\begin{equation}
\begin{aligned}
 \d \bar{g}_Z &= -\frac{1}{\sqrt2} \, \d G_F-\frac{1}{2}\frac{\d m_Z^2}{\hat{m}_Z^2}+\frac{s_{\hat\theta} c_{\hat\theta}}{\sqrt2 \hat{G}_F}C_{HWB},\\
 &=-\frac{1}{4\sqrt2 \hat{G}_F}\left( C_{HD}+4 \, C_{H\ell}^{(3)}-2 C_{ll}\right),
 \end{aligned}
\end{equation} 
and it is unchanged moving between the $\{\hat{m}_W, \hat{m}_Z,\hat{G}_F\}$ and $\{\a, \hat{m}_Z,\hat{G}_F\}$ schemes.
The couplings $g_A^f$ and $g_V^\nu$ are also unchanged moving between these schemes. \\

{\bf \noindent Effective $W^\pm$ couplings:}  For the coupling of the $W^\pm$ boson we define
\begin{equation}\label{Wcouplings.def}
 \LL_{W,eff} =  - 2^{3/4} \hat{m}_W \sqrt{\hat{G}_F}  W^+_\mu\left[\bar{\nu}\g^\mu \left(g_V^{W_\pm,\ell}-g_A^{W_\pm,\ell}\g_5\right)e+
 \bar{u}\g^\mu \left(g_V^{W_\pm,q}-g_A^{W_\pm,q}\g_5\right)d\right]+\text{h.c.}\,,
\end{equation} 
with $g_{V/A}^{W_\pm,\ell} = (g_{V/A}^{W_\pm,\ell})_{SM} + \delta (g_{V/A}^{W_\pm,\ell})$ and $(g_{V/A}^{W_\pm,\ell})_{SM}= 1/2$ while
\begin{align} 
 \d (g_V^{W_\pm,\ell})= \d (g_A^{W_\pm,\ell}) &= \frac{1}{2 \, \sqrt{2} \, \hat{G}_F} \, C_{H\ell}^{(3)} - \frac{ \delta G_F}{2 \, \sqrt{2}}, \\
 \d (g_V^{W_\pm,q})= \d (g_A^{W_\pm,q}) &= \frac{1}{2 \, \sqrt{2} \, \hat{G}_F} \, C_{H q}^{(3)} - \frac{ \delta G_F}{2 \, \sqrt{2}}.
\end{align}

 {\bf \noindent Effective photon couplings:} 
For the effective coupling of the photon we define
\bea
 \LL_{A,eff} = - \hat{e} \,  \left[Q_x (1+ \delta e/ \hat{e}) \, J_{\mu}^{A, x}  \right] A^{\mu}.
 \eea
and $Q_x= \{2/3,-1/3, -1 \}$ for $x = \{u,d,\ell\}$. Where the effective coupling in the canonically normalized SMEFT \cite{Alonso:2013hga,Berthier:2015oma}
expressed in this set of input observables is
\bea
\frac{\delta e}{\hat{e}} \equiv \frac{\delta \alpha}{2 \, \hat{\alpha}} = -\frac{\delta G_F}{\sqrt{2}} + \dfrac{\d m_Z^2}{\hat{m}_Z^2} \frac{\hat{m}_W^2}{2 \, (\hat{m}_W^2 - \hat{m}_Z^2)}
- \frac{C_{HWB}}{\sqrt{2} \, \hat{G}_F} \frac{\hat{m}_W}{\hat{m}_Z} \, s_{\hat\theta}.
\eea
The observability of shifts in the effective photon couplings requires a measurement in addition to the near $Z$-pole LEP measurements to constrain all parameters in the SMEFT.
In the fit results we report below, we use $e^+ e^- \rightarrow e^+ e^- $ scattering for this purpose. \\

{\bf \noindent Triple gauge boson interaction effective couplings:}
We use the parameterization of the $\rm C$ and $\rm P$ even Effective TGC Lagrangian \cite{Hagiwara:1986vm}
\begin{equation}
\begin{aligned}\label{effectivetgc}
\frac{\LL_{WWV,eff}}{-i \, \hat{g}_{WWV}} =& 
g_1^V \Big( W^+_{\mu\nu} W^{- \mu} V^{\nu} - W^+_{\mu} V_{\nu} W^{- \mu\nu} \Big) + \kappa_V W_\mu^+ W_\nu^- V^{\mu\nu} +\frac{i\lambda_V}{\hat{m}_W^2}V^{\mu\nu}W^{+\rho}_\nu W^-_{\rho\mu},
\end{aligned} 
\end{equation} 
where $V=\{Z,\g\}$ while $V_{\mu \nu} = \partial_{\mu} V_{\nu} - \partial_{\nu} V_{\mu}$ and $W^{\pm}_{\mu \nu}=\partial_{\mu} W^{\pm}_{\nu} - \partial_{\nu}W^{\pm}_{\mu}$. 
The couplings are defined as $\hat{g}_{WWZ} = \hat{e} \, \cot \hat{\theta}$, $\hat{g}_{WW \gamma} = \hat{e}$,
$\kappa_V = 1 + \delta \kappa_V$, $\lambda_V = \delta \lambda_V$ and $g_1^V = 1+ \d g_1^V$.
In the $\{\hat{m}_W, \hat{m}_Z,\hat{G}_F\}$ scheme one finds
\begin{align}
 \d g_1^\gamma &= \frac{1}{4\sqrt2 \hat{G}_F}\left(C_{HD}\frac{\hat{m}_W^2}{\hat{m}_W^2-\hat{m}_Z^2}-4C_{H\ell}^{(3)}+2C_{ll}-C_{HWB}\frac{4\hat{m}_W}{\sqrt{\hat{m}_Z^2-\hat{m}_W^2}}\right), \\
 \d g_1^Z &=\frac{1}{4\sqrt2 \hat{G}_F}\left(C_{HD}-4C_{H\ell}^{(3)}+2 C_{ll}+4\frac{\hat{m}_Z}{\hat{m}_W}\sqrt{1-\frac{\hat{m}_W^2}{\hat{m}_Z^2}}C_{HWB}\right),\\
 \d \kappa_\gamma &=  \frac{1}{4\sqrt2 \hat{G}_F}\left(C_{HD}\frac{\hat{m}_W^2}{\hat{m}_W^2-\hat{m}_Z^2}-4C_{H\ell}^{(3)}+2C_{ll}\right),\\
 \d \kappa_Z &= \frac{1}{4\sqrt2 \hat{G}_F}\left(C_{HD}-4C_{H\ell}^{(3)}+2C_{ll} \right),\\
 \d \lambda_\gamma &=6 \, s_{\hat{\theta}} \, \frac{\hat{m}^2_W}{\hat{g}_{WWA}} \,  C_W,\\
 \d \lambda_Z &=6 \, c_{\hat{\theta}} \, \frac{\hat{m}^2_W}{\hat{g}_{WWZ}} \,  C_W.
\end{align}
The $\d \kappa_Z = \d g_1^Z - t_{\theta}^2 \, \d \kappa_\gamma$ relationship identified holds in the SMEFT Lagrangian with $\mathcal{L}_6$ corrections in the $\{\a, \hat{m}_Z,\hat{G}_F\}$-scheme, but is not satisfied when including $\mathcal{L}_8$ corrections \cite{Hagiwara:1993ck}. This relation is not satisfied in the $\{\hat{m}_W, \hat{m}_Z,\hat{G}_F\}$ scheme, even considering $\mathcal{L}_6$ corrections, however a more general relation
\bea\label{generalTGCrelation}
\d \kappa_Z - \d g_1^Z  = - t_{\theta}^2 (\d \kappa_\gamma - \d g_1^\gamma),
\eea
holds in both schemes considering $\mathcal{L}_6$ corrections. The reason this relation holds is effective TGC corrections come about in two
ways considering $\mathcal{L}_6$ corrections. Effective shifts introduced due to relating SM couplings to input parameters and direct contributions to anomalous TGC couplings (not of a $\lambda_V$ form). The former class of corrections come in a form that respects $\d \kappa_V - \d g_1^V =0$.
The later set of corrections at $\mathcal{L}_6$ comes about due to $C_{HWB}$ in either input parameter set, which respects Eqn.~\ref{generalTGCrelation}.
The relation $\delta \lambda_\gamma = \delta \lambda_Z$ holds in both input parameter sets. 

\subsection{\texorpdfstring{$\{\hat{m}_W, \hat{m}_Z,\hat{G}_F\}$}{\{mW,mZ,GF\}} scheme benefits}\label{sec:mw_benefits}
The $\{\hat{m}_W, \hat{m}_Z,\hat{G}_F\}$ input parameter scheme has been in common use in the SM precision calculating community \cite{Sirlin:1980nh,Denner:1991kt,Freitas:2016sty} but this scheme has not been considered extensively in previous studies in the SMEFT.\footnote{A notable set of exceptions to this statement are in Refs.~\cite{Passarino:2012cb,Ghezzi:2015vva,Gauld:2015lmb,Gauld:2016kuu,Passarino:2016pzb}.} This is an unfortunate historical accident due to the precise measurement
of $\hat{m}_W$ at the Tevatron appearing after LEP data. The demonstration of the robustness of such transverse variable measurements of $\hat{m}_W$ against measurement bias in the SMEFT \cite{Bjorn:2016zlr}, and the precise measurements starting to appear from LHC \cite{Aaboud:2017svj} indicates that using this scheme is numerically sound in studies of this form and can have a number of benefits.

A key benefit is related to lifting the reparameterization invariance present in $\bar{\psi} \psi \rightarrow \bar{\psi} \psi$ scattering in global data analyses in a consistent fashion. When the $\{\hat{\alpha}, \hat{m}_Z,\hat{G}_F\}$ input scheme is used
and $\bar{\psi} \psi \rightarrow \bar{\psi} \psi \, \bar{\psi} \psi$ observables are employed for this purpose, 
a problem at leading order is introduced due to the need to expand the pole of the $W^\pm$ boson propagators in SMEFT corrections. To perform a $\chi^2$ fit the expansion
\bea
\bar{\chi}\left(s_{ij}\right)
&=& \frac{1}{\left(s_{ij} - \bar{m}_W^2 \right)^2 + \left(\bar{\Gamma}_W \bar{m}_W\right)^2} = \frac{1}{\left(s_{ij} -\hat{m}_W^2 \right)^2 + \left(\hat{\Gamma}_W \hat{m}_W\right)^2} \left[1 + \delta \chi \left(s_{ij}\right)\right], \nonumber 
\eea
is made, where the propagator modification is given by \cite{Berthier:2016tkq}
\begin{align}
    \delta \chi \left(s_{ij}\right) &= \frac{\left[- 2\left(s_{ij} - \hat{m}_W^2\right) + \Gamma_W^2 \right]\delta m_W^2 - 2 \Gamma_W \hat{m}_W^2 \delta \Gamma_W}{\left(s_{ij} - \hat{m}_W^2 \right)^2 + \left(\hat{m}_W \hat{\Gamma}_W\right)^2}. \nonumber
\end{align}
Here the bar superscript indicates a parameter at tree level in the canonically normalized SMEFT, $\delta X$ indicates the complete correction to the quantity $X$
due to $\mathcal{L}_6$ corrections, and the hat superscript notation indicates a measured parameter. $s_{ij} = (p_i+ p_j)^2$ for the four momentum $p_{i,j}$ carried by the final states.
The shift in the $W^\pm$ mass pole in the $\{\hat{\alpha}, \hat{m}_Z,\hat{G}_F\}$ scheme is the same order as
the SMEFT corrections. This formally introduces an ambiguity into the global constraint picture of the Wilson coefficient the same order as the Wilson coefficients fit to,
as the requirement to expand around the physical poles to obtain a gauge invariant decomposition of the total cross section \cite{Veltman:1963th,Stuart:1991xk,Grunewald:2000ju}
is violated.
Using a $\{\hat{m}_W, \hat{m}_Z,\hat{G}_F\}$ input scheme avoids this shift in the pole mass in an analysis of $\bar{\psi} \psi \rightarrow \bar{\psi} \psi \, \bar{\psi} \psi$
observables. 

Another benefit of the $\{\hat{m}_W, \hat{m}_Z,\hat{G}_F\}$-input scheme is the one loop corrections in this scheme are arguably
easier to implement \cite{Gauld:2015lmb,Gauld:2016kuu}. Finally, the measurement scales of the input parameters are closer together, minimizing large logs in the
perturbative expansion of observables.

\subsubsection{\texorpdfstring{$\{\hat{m}_W, \hat{m}_Z,\hat{G}_F\}$}{\{mW,mZ,GF\}} numerical predictions for LEP observables}

Predictions for the observables used in the global data analyses reported in Refs.~\cite{deBlas:2014ula,Falkowski:2014tna,Buckley:2015lku,Berthier:2015oma,Berthier:2015gja,Berthier:2016tkq,deBlas:2016ojx,Butter:2016cvz} use the $\{\hat{\alpha}, \hat{m}_Z,\hat{G}_F\}$ input parameter set.
For many collider observables, the theoretical and experimental error assigned to a SM prediction is far larger than the scheme dependence of an observable.
In this case theory predictions being reformulated switching between input parameter schemes will have small numerical effects on constraints. 
However, a subset of the LEPI pseudo-observables are a special case of precision in experimental and theoretical prediction, rising to $\sim 0.1 \%$ level precision in a few cases ($\bar{R}_\ell$, $\bar{\sigma}_h$, $\bar{\Gamma}_Z$), and should be reformulated switching between schemes. 

\begin{figure}
  \centering
   \includegraphics[width=0.495\textwidth]{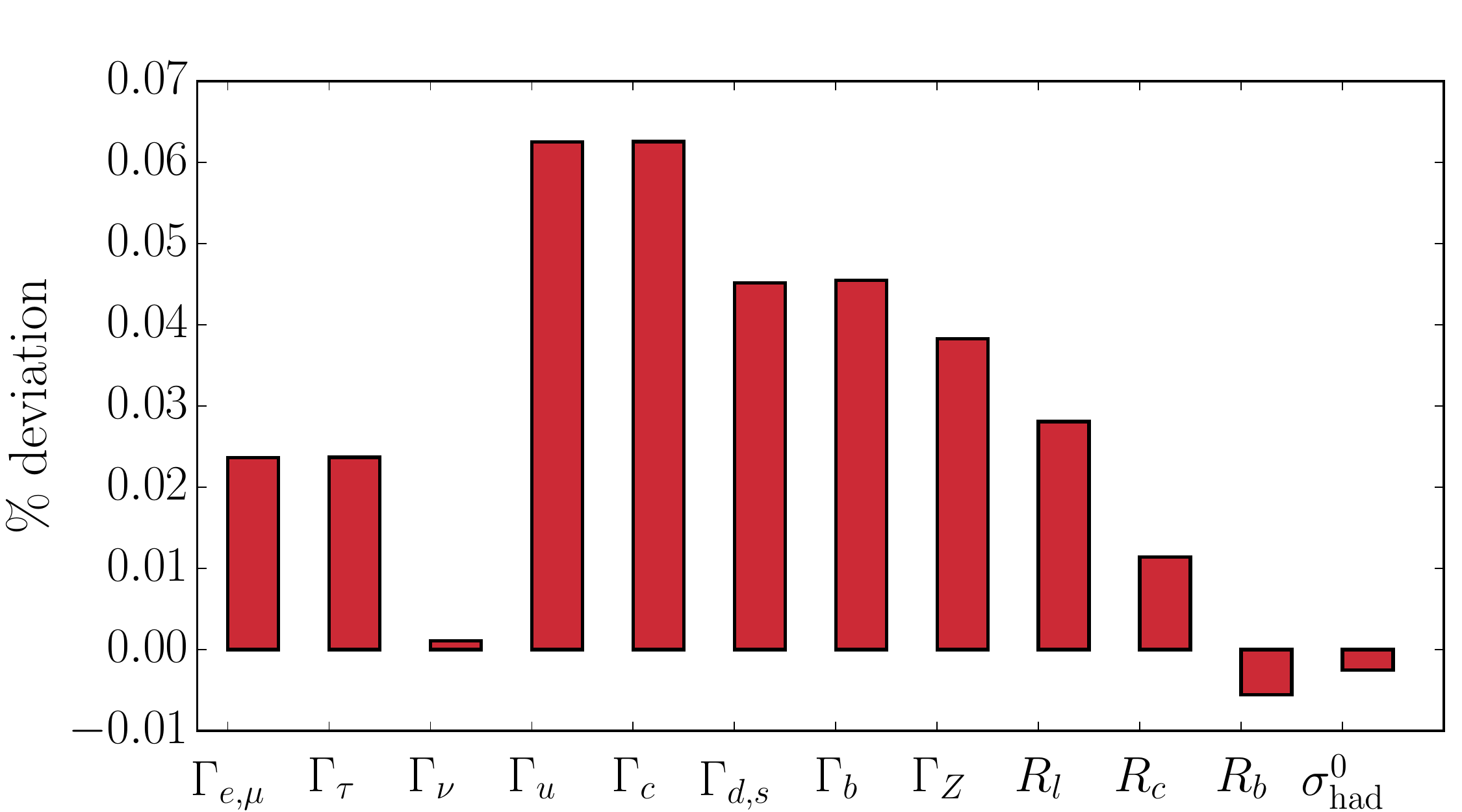}
     \includegraphics[width=0.495\textwidth]{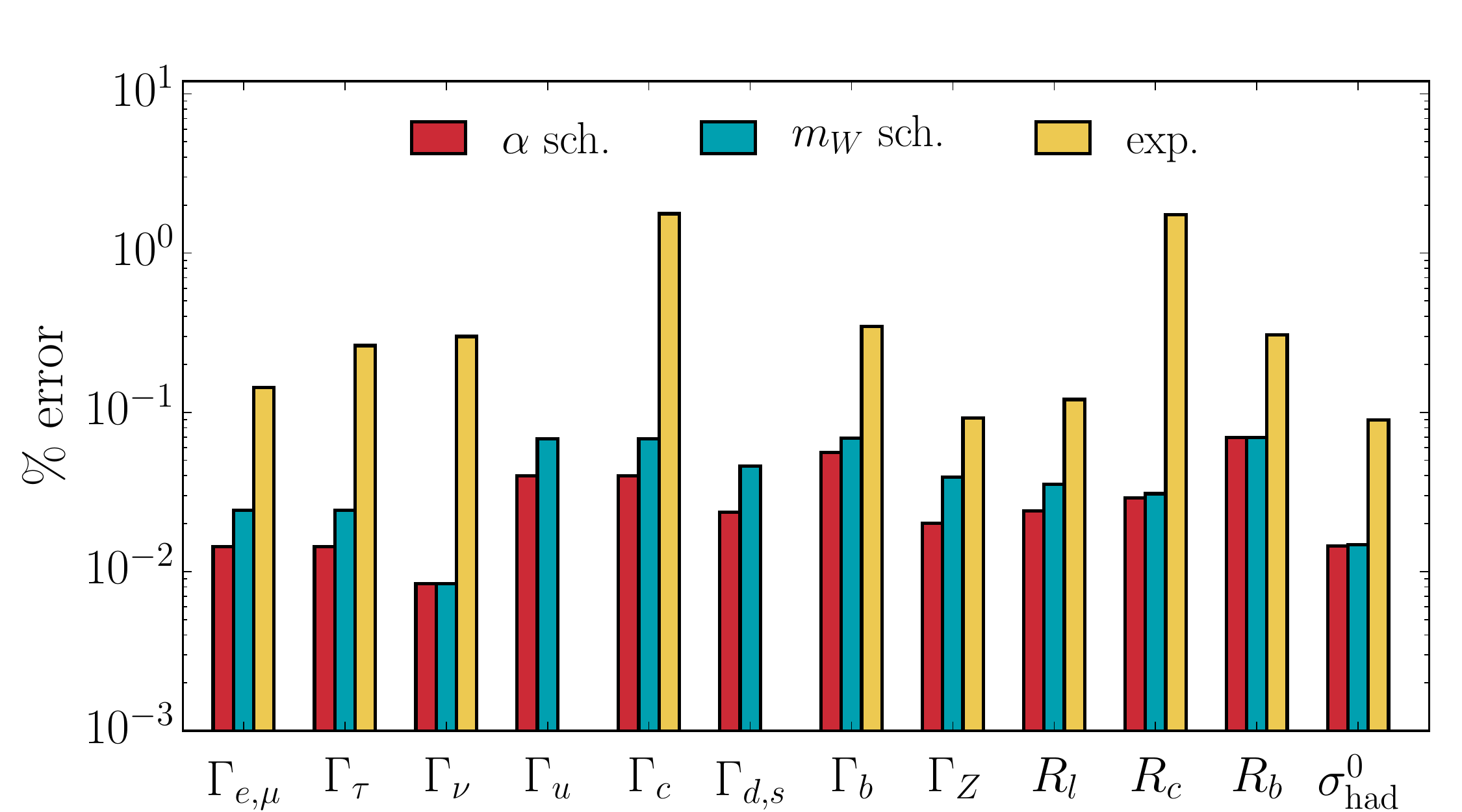}
  \caption{The left figure shows the relative $\%$ level change in each observable shifting from the $\{\hat{\alpha}, \hat{m}_Z,\hat{G}_F\}$ to the $\{\hat{m}_W, \hat{m}_Z,\hat{G}_F\}$
input scheme defined as $(X_{m_W}-X_{\alpha})/X_{\alpha} \times 100 \%$. The right handed figure shows
the total theoretical error in each observable in the $\{\hat{\alpha}, \hat{m}_Z,\hat{G}_F\}$ scheme (left handed red bar), the total theoretical error in each observable in the $\{\hat{m}_W, \hat{m}_Z,\hat{G}_F\}$ scheme (middle blue bar) and the experimental error (right yellow bar) when quoted in Ref.~\cite{ALEPH:2005ab}. 
The numerical results are reported in Table~\ref{schemeresults}.}
  \label{Fig:schemeshift}
\end{figure}

\begin{center}
\begin{table}
\centering
\tabcolsep 8pt
\begin{tabular}{|>{$}c<{$}c|*3{r@{ $\pm$ }l|}}
\toprule
\multicolumn{2}{|c|}{Observable} & \multicolumn{2}{c|}{$\{\hat{\alpha}, \hat{m}_Z,\hat{G}_F\}$ inputs} & \multicolumn{2}{c|}{$\{\hat{m}_W, \hat{m}_Z,\hat{G}_F\}$ inputs} & \multicolumn{2}{c|}{Exp. result \cite{ALEPH:2005ab}\footnote{Specifically these results are taken from Tables 7.1 and 8.4 of Ref.~\cite{ALEPH:2005ab}.}} \\ \midrule
\Gamma_{e,\mu}& [\rm MeV] & 83.966 & 0.012  & 83.986 & 0.020 & 83.92 & 0.12 \\
\Gamma_{\tau}& [\rm MeV] & 83.776 & 0.012  & 83.796 & 0.020  & 84.08 & 0.22 \\
\Gamma_{\nu}&  [\rm MeV] & 167.156 & 0.014  & 167.158 & 0.014 & 166.333 & 0.5 \\
\Gamma_{u}& [\rm MeV]& 299.95 & 0.12 & 300.149 & 0.20  & \multicolumn{2}{c|}{-}\\
\Gamma_{c}&  [\rm MeV] & 299.87 & 0.12  & 300.07 & 0.20 & 300.5 & 5.3 \\
\Gamma_{d,s}&  [\rm MeV] & 382.78 & 0.09  & 382.96 & 0.18 & \multicolumn{2}{c|}{-}\\
\Gamma_{b}&  [\rm MeV] & 375.73 & 0.21  & 375.91 & 0.26 & 377.6 & 1.3 \\
\Gamma_{Z} & [\rm MeV] & 2494.3 & 0.5  & 2495.3 & 1.0 & 2495.2 & 2.3 \\
R_{\ell} && 		20.752 & 0.005  & 20.758 & 0.007 & 20.767 & 0.025 \\
R_{c} &&		 0.17223 & 0.00005  & 0.172254 & 0.000053 & 0.1721 & 0.003 \\
R_{b} &&		 0.2158 & 0.00015  & 0.21579 & 0.00015 & 0.21619 & 0.00066 \\
\sigma^0_{\rm Had}& [\rm pb]& 41488 & 6  & 41486.5 & 6.1 & 41541 & 37\\
\bottomrule
\end{tabular}
\caption{Predictions for LEPI observables in the two input parameter schemes.\label{schemeresults}}
\end{table}
\end{center}
\vspace{-1cm}
Predictions of the LEPI pseudo-observables in the $\{\hat{m}_W, \hat{m}_Z,\hat{G}_F\}$-input parameter scheme are produced as follows.\footnote{We thank A. Freitas
for suggesting this approach.} We use the expansion formula reported in Ref.~\cite{Freitas:2014hra} for the LEPI pseudo-observables as a function of $\{\hat{m}_h,\hat{m}_Z,\hat{m}_t,
\Delta \hat{\alpha}, \hat{\alpha}(\hat{M}_z) \}$
combined with the expansion formuli reported in Ref.~\cite{Awramik:2003rn} for $\hat{m}_W$ as a function of the same set of inputs. We solve
the latter for $\hat{m}_W$ to replace dependence on $\Delta \hat{\alpha}$ in Ref.~\cite{Freitas:2014hra}
in favour of $\hat{m}_W$. We use the quoted value of the Tevatron average measurement of $\hat{m}_W = 80.387 \pm 0.016 \, {\rm GeV}$
to then produce effective predictions of the  LEPI pseudo-observables as a function of $\{\hat{m}_W, \hat{m}_Z,\hat{G}_F\}$.
Using this method we find the results in Table \ref{schemeresults}.
The observables reported are defined as 
\begin{align}
\bar{\Gamma}_i &= \frac{\sqrt{2} \, \hat{G}_F \, \hat{m}_Z^3 \, N_c}{3 \, \pi} \left(|\bar{g}_V^i|^2 + |\bar{g}_A^i|^2 \right),& \quad 
\bar{\Gamma}_{had} &= \bar{\Gamma}_{u} + \bar{\Gamma}_{d} +\bar{\Gamma}_{c} + \bar{\Gamma}_{s} + \bar{\Gamma}_{b}, \\
\bar{R}_{c,b} &= \frac{\bar{\Gamma}_{c,b}}{\bar{\Gamma}_{had}}, & \quad
\bar{R}_{\ell} &= \frac{\bar{\Gamma}_{had}}{\bar{\Gamma}_{\ell}}, \\
\bar{\sigma}_{had}^{0} &= \frac{12 \, \pi}{\bar{m}_Z^2} \, \frac{\bar{\Gamma}_e \, \bar{\Gamma}_{had}}{\bar{\Gamma}_Z^2}.
\end{align}
 The impact of the change between these schemes is illustrated in Fig.~\ref{Fig:schemeshift}. To shift to the $\{\hat{m}_W, \hat{m}_Z,\hat{G}_F\}$
 scheme we also introduce a theoretical error for the $m_W$ mass. We use the inferred dependence in the expansion formuli of Ref.~\cite{Freitas:2014hra,Awramik:2003rn} on $\hat{m}_{W}$, defining this error for each observable in the scheme $x_i = \{m_W, \alpha\}$ as $(\nabla X)_{x_i}$ where
 \bea
(\nabla X)_{m_W} = \sqrt{(\nabla X)_{\alpha}^2 + \left(\abs{\frac{\partial X}{\partial m_W}} (\nabla \, m_W)\right)^2},
 \eea
and $ (\nabla \, m_W) = 0.016$~GeV. This study should be supplemented with a dedicated
analysis producing predictions for the full set of EWPD observables directly, without use of the intermediate expansion formulas in Ref.~\cite{Awramik:2003rn,Freitas:2014hra}. However, as the theoretical error in both schemes are below the experimental errors in all cases, this initial study is sufficient for our purpose.

\subsection{Numerical global fit results}\label{fitresults}
A global fit analysis to LEP data using the $\{\hat{\alpha}, \hat{m}_Z,\hat{G}_F\}$ and the $\{\hat{m}_W, \hat{m}_Z,\hat{G}_F\}$ schemes is used here to quantify the impact of the inputs choice on the resulting constraints on the Wilson coefficients.
This analysis is presented in two consecutive stages: in a first step only 31 LEPI observables, obtained from measurements of $\bar{\psi} \psi \rightarrow \bar{\psi} \psi$ scattering processes, are included. In both schemes
the results obtained exhibit two unconstrained directions. 
As a second step, LEPII measurements of $\bar{\psi} \psi \rightarrow \bar{\psi} \psi \bar{\psi} \psi$ scattering through $W^\pm$ currents are incorporated in the fit, in order to lift the unconstrained directions. 
\\
\\
{\bf Fit methodology:} We employ the fit method of Refs.~\cite{Berthier:2015gja,Berthier:2016tkq}. The measured value of a given observable $\hat O_i$ is assumed to be a gaussian variable centered about the theoretical prediction in the SMEFT $\bar O_i$ so that the likelihood function can be defined as
\begin{equation}
 L(C) = \frac{1}{\sqrt{(2\pi)^n\det(V)}}\exp\left(-\frac{1}{2}(\hat O- \bar O)^T V^{-1} (\hat O - \bar O)\right),
\end{equation}
where the $n$ dimensional vectors $\hat O = (\hat O_1,\dots, \hat O_n)$, $\bar O = (\bar O_1,\dots, \bar O_n)$ have been introduced and $V$ represents the covariance matrix
\begin{equation}
 V_{ij}=\Delta_i^{exp}\rho_{ij}^{exp}\Delta_j^{exp} + \Delta_i^{th}\rho_{ij}^{th}\Delta_j^{th}.
\end{equation}
Here $\rho^{exp}/\rho^{th}$ are the experimental/theoretical correlation matrices and $\Delta^{exp}_i/\Delta^{th}_i$ are the experimental/theoretical error of the observable $O_i$. The theoretical error for each observable is defined so as to contain both the SM theoretical uncertainty $\Delta_{i,{\rm SM}}$ and a constant relative SMEFT theory error $\Delta_{\rm SMEFT}$~\cite{Berthier:2015oma} defined as
\begin{equation}
 \Delta^{th}_i=\sqrt{\Delta_{i,{\rm SM}}^2+\Delta_{\rm SMEFT}^2 \bar O_i^2}.
\end{equation}
We define the $\chi^2$ variable as $\chi^2=-2\log(L(C))$. 
Potential unconstrained directions in the analysis can finally be identified as the null eigenvectors of the Fisher information matrix
\begin{equation}
 \mathcal{I}_{ij} = \frac12\left(\frac{\partial^2}{\partial C_i \,\partial C_j}\chi^2\right).
\end{equation}

\subsubsection{LEPI observables}
The first stage of the global analysis follows closely the procedure presented in Refs.~\cite{Berthier:2015oma,Berthier:2015gja}, the main difference being the fact that 
we use  31 observables measured at LEPI instead of the 103 observables considered in Refs.~\cite{Berthier:2015oma,Berthier:2015gja}. This choice does not limit the power of the fit in a significant way and it suffices to illustrate the main physical conclusions. We include measurements of
\begin{itemize}
 \item the near $Z$-pole observables listed in Table~\ref{schemeresults} and the $W^\pm$ mass,
 \item the forward-backward asymmetries $\mathcal{A}_{FB}^{0, f}$ for $f=\{c, b,\ell\}$,
 \item the differential distributions of bhabha scattering $d\sigma(e^+ e^-\to e^+e^-)/d\cos\theta$.
\end{itemize}
Notice that the measurement of the $W^\pm$ mass represents a constraint only when the $\{\hat{\alpha}, \hat{m}_Z,\hat{G}_F\}$ scheme is adopted, while the inclusion of $e^+ e^-\to e^+e^-$ scattering data is required in order to introduce an independent constraint on the value of $\hat\alpha$ in the $\{\hat{m}_W, \hat{m}_Z,\hat{G}_F\}$ scheme.

The theoretical SM predictions in  the $\{\hat{m}_W, \hat{m}_Z,\hat{G}_F\}$ input scheme for the first category of observables were computed in the previous section, and the results are listed in Table~\ref{schemeresults}.
The theoretical values for the latter two categories, instead, vary by a quantity smaller than the theoretical error when switching between input parameter schemes. As such the theory predictions were taken to be the same as the values quoted in Ref.~\cite{Berthier:2015gja} which also lists the experimental data used and errors.

The analytic dependence of the observables on the Lagrangian parameters was given in Ref.~\cite{Berthier:2015oma} and is formally unchanged in the $\{\hat{m}_W, \hat{m}_Z,\hat{G}_F\}$ scheme. The main difference with the $\{\hat{\alpha}, \hat{m}_Z,\hat{G}_F\}$-scheme computation is the fact that $\bar{\alpha}$ and $\bar{g}_1^\gamma$ now carry a dependence on the SMEFT parameters, while $\bar{m}_W$ does not.
For this limited set of observables, the subset of relevant $\mathcal{L}_6$ Wilson coefficients are
\begin{equation}
\tilde{C}_i \equiv \frac{\bar{v}_T^2}{\Lambda^2} \{
 C_{He}, C_{Hu}, C_{Hd}, \CHls, \CHlt, \CHqs, \CHqt, C_{HWB}, C_{HD}, C_{ll}, C_{ee}, C_{le}
 \}.
\end{equation}
Using the $\{\hat{\alpha}, \hat{m}_Z,\hat{G}_F\}$ input scheme and normalizing to the coefficient of $C_{He}$ the null eigenvectors
of the Fisher information matrix are
\begin{align}
w^{\a}_1&= \frac{\bar{v}_T^2}{\Lambda^2} \left(\frac{1}{3}C_{Hd}-2C_{HD}+ C_{He}+\frac{1}{2} \CHls-\frac{1}{6} \CHqs-\frac{2}{3} C_{Hu}-1.29 (\CHqt+ \CHlt)+1.64 C_{HWB}\right),\label{a1}  \\
w^{\a}_2&=\frac{\bar{v}_T^2}{\Lambda^2} \left(\frac{1}{3}C_{Hd}-2C_{HD}+ C_{He}+\frac{1}{2} \CHls-\frac{1}{6} \CHqs-\frac{2}{3} C_{Hu}+2.16 (\CHqt+ \CHlt)-0.16 C_{HWB}\right)\label{a2}. 
\end{align}
Performing the fit in the $\{\hat{m}_W, \hat{m}_Z,\hat{G}_F\}$ scheme the unconstrained directions are
\begin{align}
w^{m_W}_1&= \frac{\bar{v}_T^2}{\Lambda^2}\left(\frac{1}{3}C_{Hd}-2C_{HD}+ C_{He}+\frac{1}{2} \CHls-\frac{1}{6} \CHqs-\frac{2}{3} C_{Hu}-1.24 (\CHqt+ \CHlt)+1.60 C_{HWB}\right),\label{mw1}  \\
w^{m_W}_2&= \frac{\bar{v}_T^2}{\Lambda^2}\left(\frac{1}{3}C_{Hd}-2C_{HD}+ C_{He}+\frac{1}{2} \CHls-\frac{1}{6} \CHqs-\frac{2}{3} C_{Hu}+ 2.20 (\CHqt+ \CHlt)- 0.24 C_{HWB}\right)\label{mw2}.  
\end{align}

Since all the observables included are extracted from measurements of  $\bar{\psi} \psi \rightarrow \bar{\psi} \psi$ processes, they satisfy the reparameterization invariance presented in Section~\ref{sec:EOMreparam}. As a consequence these unconstrained directions must be a linear combination of the vectors $w_{B,W}$ defined in Eqs.~\ref{wBW} if the reparameterization invariance identified is scheme independent. We find this is the case and the unconstrained directions decompose as 
\begin{align}
w^{\a}_1&= -w_B - 2.59 \, w_W&
w^{\a}_2&= -w_B + 4.31 \, w_W,\\
w^{m_W}_1&= -w_B - 2.48 \, w_W&
w^{m_W}_2&= -w_B + 4.40 \, w_W.
\end{align}

\begin{table}[t]
\hspace*{-1cm}
 \begin{tabular}{c|*{13}{c}}  
 \toprule
$\sqrt{s}$&  $\frac{\delta\Gamma_W}{\Gamma_W}$& $\delta g_W^\nu$& $\delta g_W^\pm$& $\delta g_V^Z$& $\delta g_A^Z$& $\delta g_1^Z$& $\delta\kappa_\gamma$& $\delta\kappa_Z$& $\delta \lambda_\gamma$& $\delta\lambda_Z$& $\frac{\delta \Gamma_Z}{\Gamma_Z}$& $\delta g_1^\gamma$& $\frac{\delta e}{e}$\\
 \midrule
188.6& 	 -17. &	 72.& 	 33.4& 	 5.72& 	 0.21&  -0.05 & -0.57 & -0.16 & -0.34 & 0.051 &	 0.0005 & -0.41& -0.98 \\
191.6& 	 -17. &  72.& 	 33.6& 	 6.26& 	 0.33&  -0.07 & -0.64 & -0.19 & -0.37 & 0.045 &	 0.0005 & -0.44& -1.08  \\
195.5& 	 -17. &  73.& 	 33.8& 	 6.91& 	 0.50&  -0.09 & -0.72 & -0.22 & -0.41 & 0.035 &	 0.0005 & -0.49&-1.20   \\
199.5& 	 -17. &  74.& 	 33.7& 	 7.52& 	 0.68&  -0.11 & -0.79 & -0.26 & -0.45 & 0.022 &	 0.0005 & -0.53& -1.33  \\
201.6& 	 -17. &  74.& 	 33.7& 	 7.82& 	 0.78&  -0.12 & -0.83 & -0.28 & -0.47 & 0.016 &	 0.0005 & -0.55& -1.39  \\
204.8& 	 -17. &	 74.& 	 33.5& 	 8.24& 	 0.93&  -0.14 & -0.89 & -0.32 & -0.47 & 0.005 &	 0.0005 & -0.58& -1.47  \\
206.5& 	 -17. &	 75.& 	 33.4& 	 8.45& 	 1.01& 	-0.15 & -0.92 & -0.33 & -0.51 & -0.001&	 0.0005 & -0.60& -1.52  \\
208.& 	 -17. &	 75.& 	 33.3& 	 8.62& 	 1.08& 	-0.16 & -0.94 & -0.35 & -0.52 & -0.007&	 0.0005 & -0.61& -1.55  \\
    \bottomrule
 \end{tabular}
\caption{Total cross section contributions (in pb) to $\bar{\psi} \psi \rightarrow \bar{\psi} \psi \bar{\psi} \psi$  production due to common shift parameters, in the $\{\hat{m}_W, \hat{m}_Z,\hat{G}_F\}$ scheme. The results are normalized for semileptonic final states: they should be multiplied for 1.01 (1/4.04) for fully hadronic (leptonic) final states. The quantity $\delta g_W^\nu=\delta g_W^\ell$ corresponds to the shift in the $W^\pm$ coupling to $e^+e^-$ in the $t$-channel diagrams, while the column $\delta g_W^\pm=\delta g_W^{q/\ell}$ accounts for the shift in each $W^\pm$ coupling to a pair of final state fermions.
The corresponding results obtained in the $\{\hat{\alpha}, \hat{m}_Z,\hat{G}_F\}$-scheme were reported in Table~2 of Ref.~\cite{Berthier:2016tkq}.}
\label{Tab:WWxs}
\end{table}
\begin{table}[t]
 \centering
 \begin{tabular}{c|*{12}{c}}  
\multicolumn{13}{c}{$\sqrt{s}=182.66$ GeV}\\
 \toprule
Bin&  $\frac{\delta\Gamma_W}{\Gamma_W}$& $\delta g_W^\nu$& $\delta g_W^\pm$& $\delta g_V^Z$& $\delta g_A^Z$& $\delta g_1^Z$& $\delta\kappa_\gamma$& $\delta\kappa_Z$& $\delta \lambda_\gamma$& $\delta\lambda_Z$& $\delta g_1^\gamma$& $\frac{\delta e}{e}$\\
 \midrule
$B_1$&	-1.5&	12.&	2.9&	4.3&	3.0&	-0.42&	-0.37&	-0.45&	-0.35&	-0.43&	-0.34&	-0.71\\
$B_2$&	-2.8&	16.&	5.4&	3.7&	2.3&	-0.29&	-0.35&	-0.38&	-0.28&	-0.32&	-0.27&	-0.62\\
$B_3$&	-5.2&	22.&	10.2&	1.7&	0.2&	-0.04&	-0.16&	-0.06&	-0.08&	0.03&	-0.12&	-0.29\\
$B_4$&	-14.1&	40.&	27.5&	-7.8&	-9.0&	1.20&    0.67&	 1.27& 	 0.68&	1.27&	 0.64&	1.30\\
   \bottomrule
 \multicolumn{13}{c}{}  \\[-3mm]
\multicolumn{13}{c}{$\sqrt{s}=205.92$ GeV}\\
 \toprule
Bin&  $\frac{\delta\Gamma_W}{\Gamma_W}$& $\delta g_W^\nu$& $\delta g_W^\pm$& $\delta g_V^Z$& $\delta g_A^Z$& $\delta g_1^Z$& $\delta\kappa_\gamma$& $\delta\kappa_Z$& $\delta \lambda_\gamma$& $\delta\lambda_Z$& $\delta g_1^\gamma$& $\frac{\delta e}{e}$\\
 \midrule
$B_1$&	-0.9&	10.&	1.8 &	4.9 &	2.9 &	-0.40&	-0.47&	-0.46&	-0.43&	-0.43&	-0.41&	-0.88\\
$B_2$&	-2.0&	15.&	4.0 &	5.1 &	2.8 &	-0.31&	-0.57&	-0.51&	-0.40&	-0.38&	-0.35&	-0.92\\
$B_3$&	-4.5&	22.&	8.8 &	3.7 &	1.2 &	-0.17&	-0.39&	-0.22&	-0.21&	-0.07&	-0.27&	-0.66\\
$B_4$&	-19.8&	59.&	39.0&	-9.5&	-11.4&	 1.48&	 0.88&	 1.63&	 0.93&	 1.67&	 0.81&	 1.69\\
\bottomrule
 \end{tabular}
\caption{Angular bin cross section contributions (in pb) to $\bar{\psi} \psi \rightarrow \bar{\psi} \psi \bar{\psi} \psi$ production in the $m_W$-input scheme due to shift parameters. The overall normalization and notation are the same as those of Table~\ref{Tab:WWxs}. The corresponding results obtained in the $\{\hat{\alpha}, \hat{m}_Z,\hat{G}_F\}$-scheme were reported in Table~3 of Ref.~\cite{Berthier:2016tkq}.}
\label{Tab:WWdiff}
\end{table}

\subsubsection{Incorporating \texorpdfstring{$\bar{\psi} \psi \rightarrow \bar{\psi} \psi \,  \bar{\psi} \psi$}{2->4} production data}
In a second stage of the analysis, LEPII measurements of $\bar{\psi} \psi \rightarrow \bar{\psi} \psi \,  \bar{\psi} \psi$ scattering via $W^\pm$ currents are incorporated in the global fit. We follow the procedure adopted in Ref.~\cite{Berthier:2016tkq}, computing the total spin-averaged cross section for the process $e^+e^-\to \bar{\psi} \psi \,  \bar{\psi} \psi$ in the SMEFT with the  $\{\hat{m}_W, \hat{m}_Z,\hat{G}_F\}$ input parameter scheme, for eight different values of the center-of-mass energy. The results are given in terms of a set of common shift parameter in Table~\ref{Tab:WWxs}.
Here the main differences with the computation in the $\{\hat{\alpha}, \hat{m}_Z,\hat{G}_F\}$-scheme are in the presence of non-vanishing contributions due to $\delta g_1^\gamma$ and $\delta e/\hat{e}\sim\delta\alpha/\hat\alpha$ and in the treatment of the pole in the $W^\pm$ propagators, which, as detailed in Sec.~\ref{sec:mw_benefits}, does not need to be expanded in this case, thus ensuring a more consistent gauge invariant decomposition of the cross section.

\begin{figure}[t]
 \centering
 \includegraphics[height=12cm, trim ={0 0 4.6cm 0}, clip]{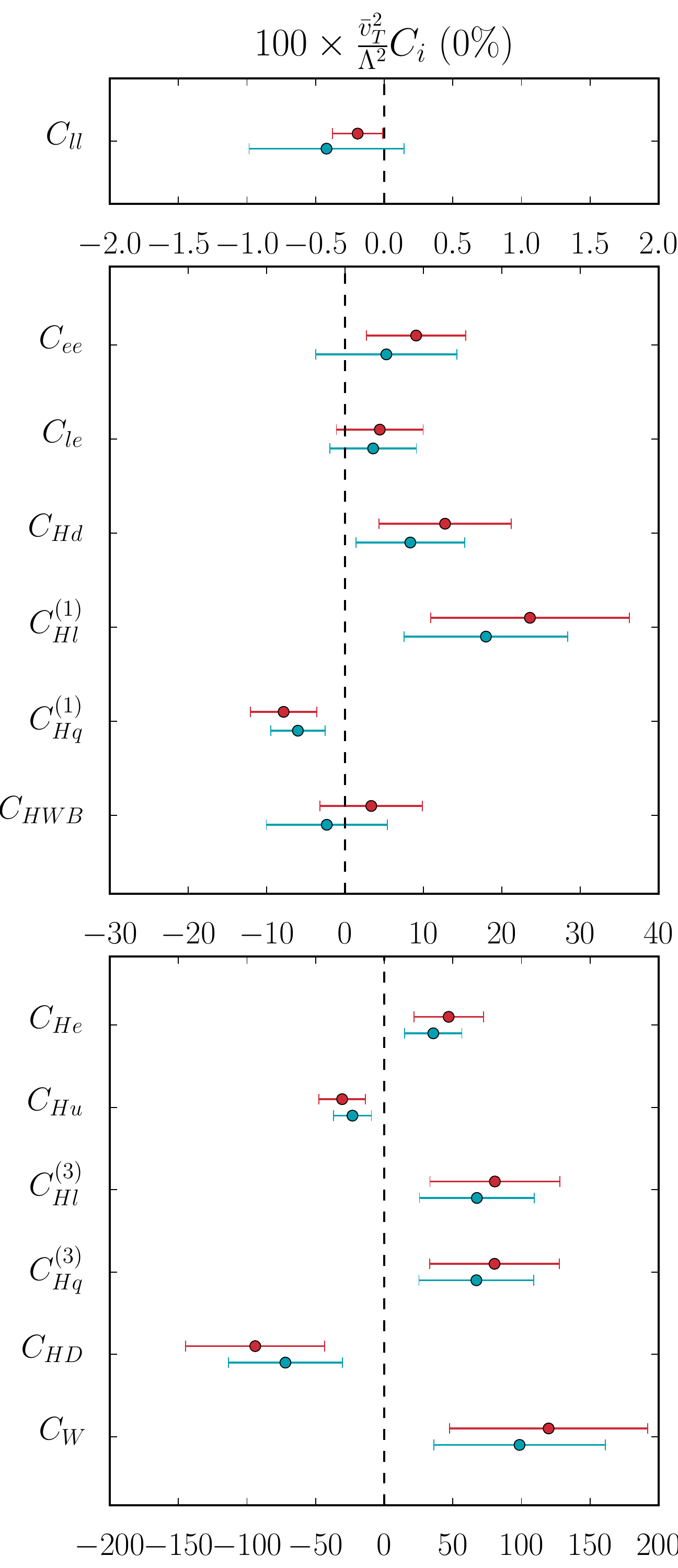}
 \includegraphics[height=12cm]{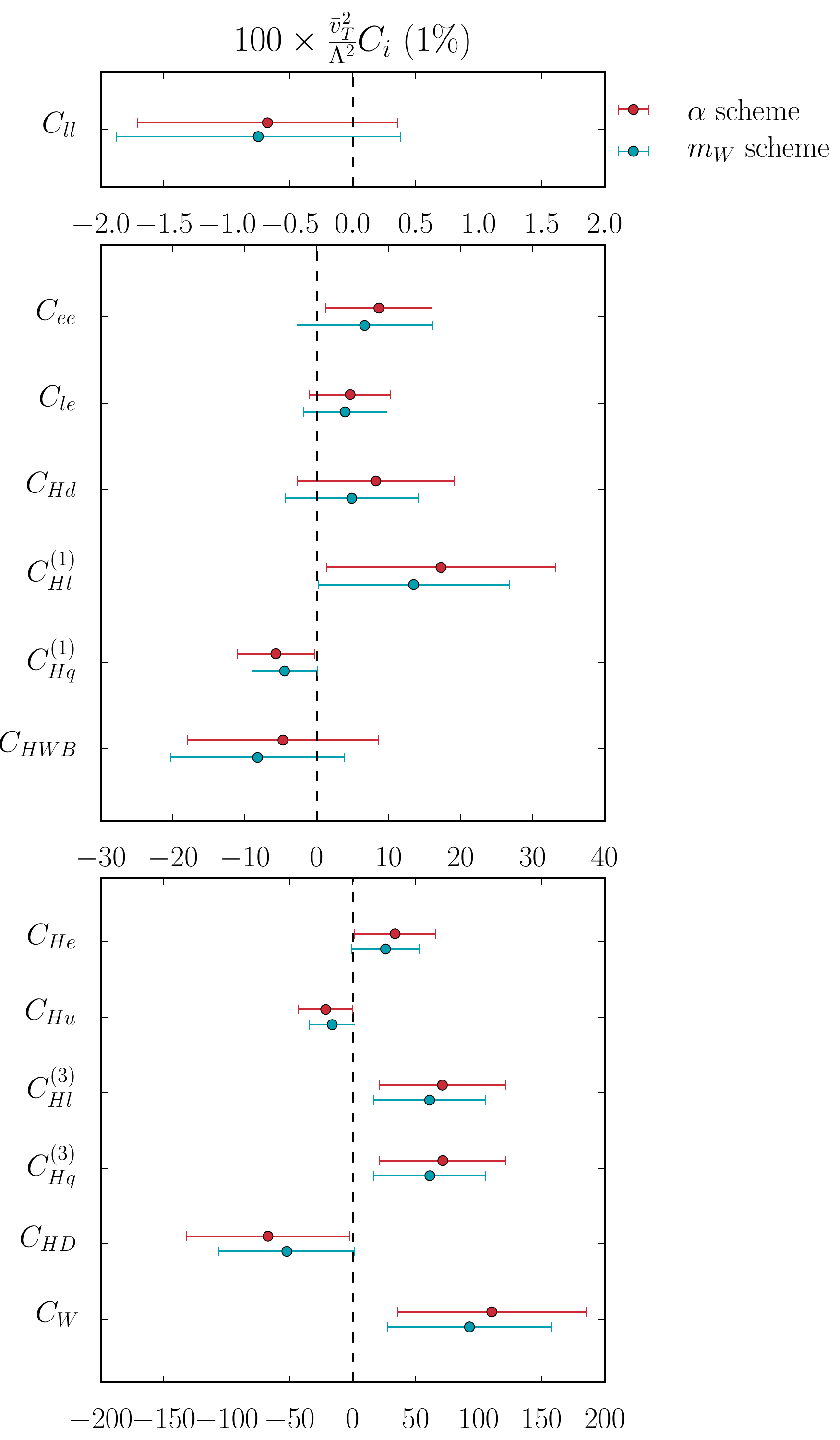}
 \caption{Best fit values of the Wilson coefficients (scaled by a factor 100) and corresponding $\pm1\sigma$ confidence regions obtained after profiling away the other parameters. Red (blue) points were obtained in the $\{\hat\alpha \,(\hat{m}_W), \hat{m}_Z,\hat{G}_F\}$ input parameter scheme. The plot to the left has been obtained assuming $\Delta_{\rm SMEFT}=0$, while the one to the right includes a theoretical error $\Delta_{\rm SMEFT}=0.01$. }
 \label{Fig:limits_profiling}
\end{figure}

We also compute the angular distribution $d\sigma/d\cos\theta$ as in Ref.~\cite{Berthier:2016tkq}, where $\theta$ is the angle formed by the momenta of the $W^+$ and of the incoming $e^-$ in the center-of-mass reference frame. In order to compare the theoretical prediction to LEPII data, we apply the kinematic cut $-0.94 < \theta_\ell < 0.94$ which ensures that, in the semileptonic final state, the angle $\theta_\ell$ between the outgoing charged lepton and the beamline does not exceed the detector acceptance of $20^o$. Finally, we compute the cross section for four bins defined by
\begin{equation}
\begin{aligned}
 B_1:& &-1 \leq& \cos\theta \leq -0.8 \qquad&
 B_2:& &-0.4 \leq& \cos\theta \leq -0.2 \\
 B_3:& &0.4 \leq& \cos\theta \leq 0.6 &
 B_4:& &0.8 \leq& \cos\theta \leq 1.
 \end{aligned}
\end{equation}
The results are given in terms of the core shift parameters in Table~\ref{Tab:WWdiff}, while the corresponding values for the $\{\hat{\alpha}, \hat{m}_Z,\hat{G}_F\}$  input choice were reported in Table~3 of Ref.~\cite{Berthier:2016tkq}.
Incorporating doubly-resonant $e^+e^-\to \bar{\psi} \psi \,  \bar{\psi} \psi$ data introduces 74 extra observables\footnote{The fit includes a total of 66 measurements of the total cross sections provided independently by the experiments L3, OPAL and ALEPH for different values of $\sqrt{s}$ and final states, plus 8 independent measurements of the angular distribution.} and an additional set of 8 relevant Wilson coefficients to the global fit:
\begin{equation}
 \tilde{C}_j = \frac{\bar{v}_T^2}{\Lambda^2}\{ 
 C_W, C_{eu}, C_{ed}, C_{lu}, C_{ld}, C_{lq}^{(1)}, C_{lq}^{(3)}, C_{eq}
 \}.
\end{equation}
Because $e^+e^-\to \bar{\psi} \psi \,  \bar{\psi} \psi$ processes are not invariant under the simultaneous rescaling of the gauge bosons fields and of their associated couplings, their inclusion in the global fit breaks the unconstrained directions in $\bar{\psi} \psi \to  \bar{\psi} \psi$ global analyses. Therefore it is possible to infer bounds on each of the 20 Wilson coefficients after profiling over the others when this data is included. These constraints are displayed in Figure~\ref{Fig:limits_profiling} for both the $\{\hat{\alpha}, \hat{m}_Z,\hat{G}_F\}$ and the $\{\hat{m}_W, \hat{m}_Z,\hat{G}_F\}$  input schemes and for two different choices of the SMEFT theoretical error due to neglected higher order effects in the analyses. See Appendix \ref{appendixresults} for the numerical results that these figures correspond to. Comparing the results of the two schemes, it is possible to notice the presence of some scheme dependence, that is comparable (but sub-dominant) to the $1 \sigma$ theory error that emerges from the fit. 
Comparing how much the bounds in the two schemes overlap when a $\sim 1\%$
SMEFT theory error is assigned, shows how considering a theoretical error for the SMEFT ameliorates the scheme dependence of global constraint results.

\begin{figure}[t]
 \centering
 \includegraphics[height=12cm, trim ={0 0 4.71cm 0}, clip]{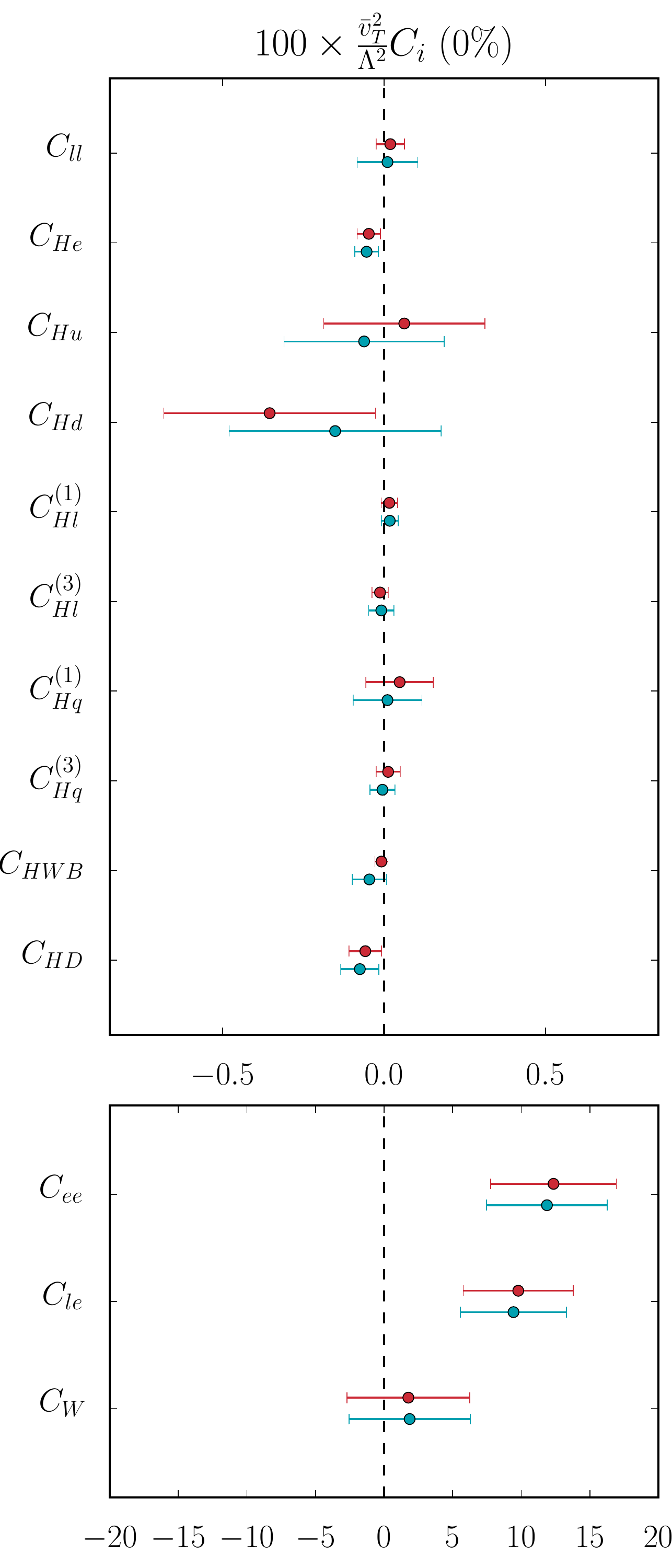}
 \includegraphics[height=12cm]{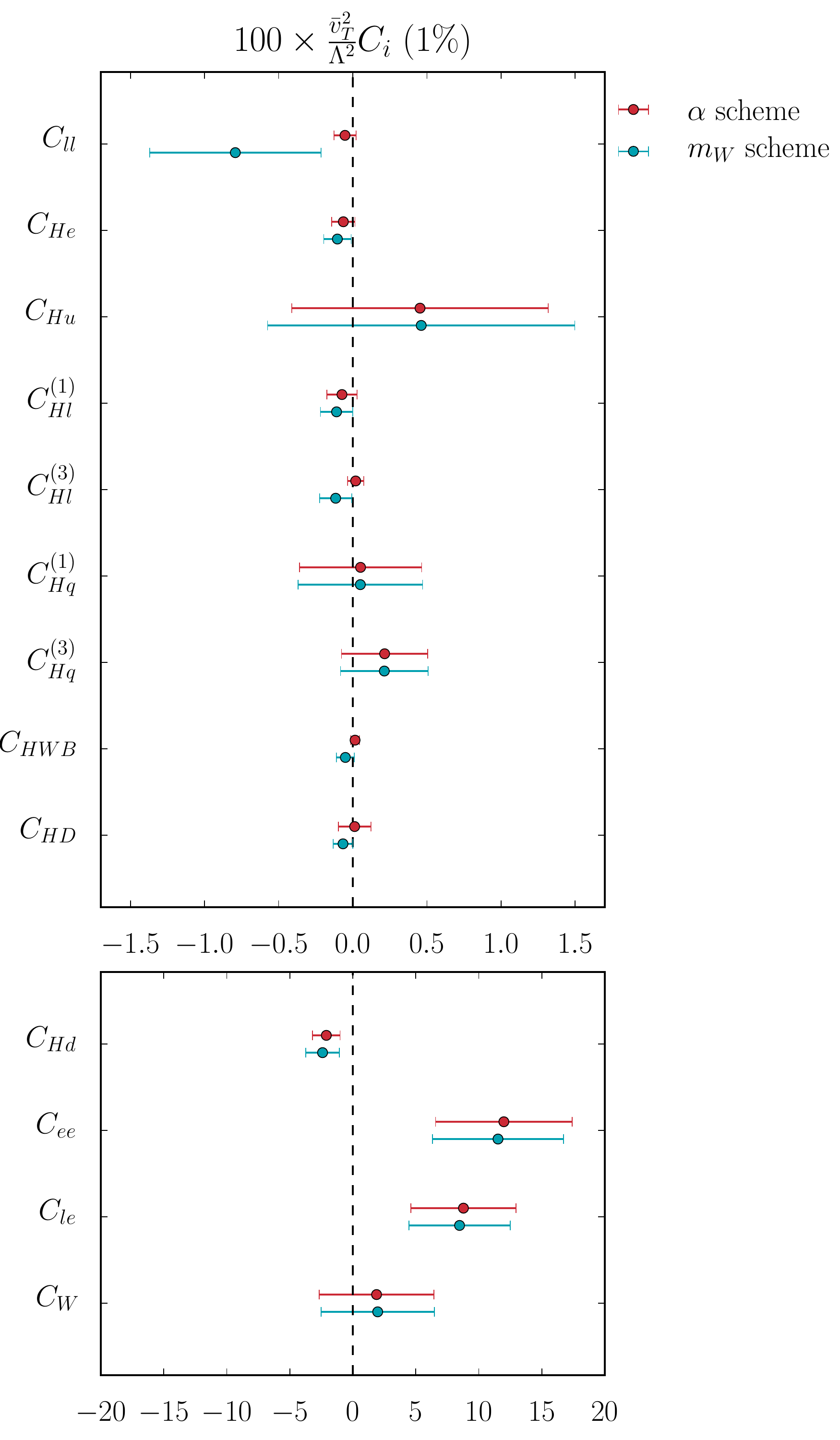}
 \caption{Best fit values of the Wilson coefficients (scaled by a factor 100) and corresponding $\pm1\sigma$ confidence regions obtained minimizing the $\Delta\chi^2$ with one parameter at a time. Red (blue) points were obtained in the $\{\hat\alpha \, (\hat{m}_W), \hat{m}_Z,\hat{G}_F\}$ input parameter scheme. The plot to the left has been obtained assuming $\Delta_{\rm SMEFT}=0$, while the one to the right includes a theoretical error $\Delta_{\rm SMEFT}=0.01$. Note that in the right plot the $x$ axis has been scaled by a factor 2 and the coefficient $C_{Hd}$ has been moved to the lower panel: increasing the theoretical error enhances the pull of the $\mathcal{A}_{FB}^{0,b}$ anomaly compared to $Z$ width data, and this relaxes by one order of magnitude the bound on this parameter.}
 \label{Fig:limits_oneatatime}
\end{figure}

We also show in Figure~\ref{Fig:limits_oneatatime}, for the sake of comparison, the constraints obtained minimizing the $\chi^2$ with one Wilson coefficient at a time. We stress that such analyses should be interpreted with significant caution, as they do not seem relatable to a consistent UV scenario inducing an operator matching pattern of this form. This is due to the non-minimal character of the SMEFT \cite{Jiang:2016czg} when the new scales introduced ($\Lambda$) have a dynamical origin.

\begin{figure}[t]
 \centering
 \includegraphics[width=\textwidth]{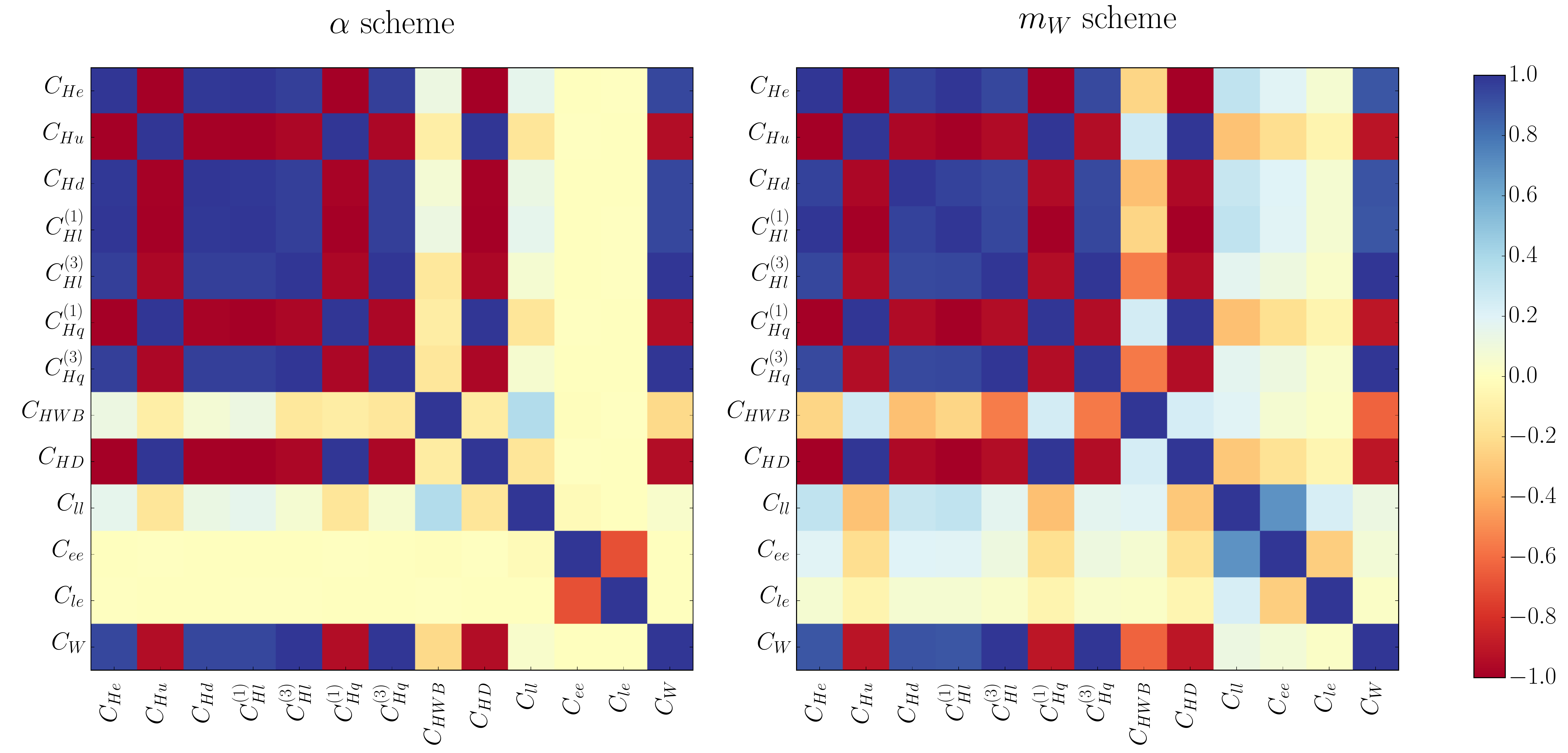}
 \caption{Color map of the correlation matrix among the Wilson coefficients, obtained assuming zero SMEFT error, for the $\{\hat{\alpha}, \hat{m}_Z,\hat{G}_F\}$ input scheme (left) and for the $\{\hat{m}_W, \hat{m}_Z,\hat{G}_F\}$ input scheme (right).}
 \label{Fig:correlation_matrices}
\end{figure}

Finally, Figure~\ref{Fig:correlation_matrices} gives a graphical representation of the correlation matrices among the Wilson coefficients obtained in both schemes. The fit space is highly correlated, irrespective of the input parameter choice. This is mostly a physical consequence of the reparameterization invariance as is demonstrated by the fact that the parameters related to the reparameterization invariance
\begin{equation}
 \{
 C_{He}, C_{Hu}, C_{Hd}, \CHls, \CHlt, \CHqs, \CHqt, C_{HD}
 \}.
\end{equation}
 are found to be strongly correlated.
The parameter $C_{HWB}$ is also involved in the unconstrained directions but its correlation is significantly washed out
by the use of $e^+e^-\to \bar{\psi} \psi \,  \bar{\psi} \psi$ processes to break the reparameterization invariance. The degree
to which $C_{HWB}$ is uncorrelated by the inclusion of this data shows significant scheme dependence, being more correlated
in the $\{\hat{m}_W, \hat{m}_Z,\hat{G}_F\}$ scheme.

\section{Conclusions}\label{conclusions}
In this paper we have explained a reparameterization invariance that is present in $\bar{\psi} \psi \rightarrow  \bar{\psi} \psi$ scattering
in the SMEFT. This invariance is broken by the inclusion of scattering data with different Feynman diagram topologies and it represents an
underlying physical reason why the fit space of the $\mathcal{L}_6$ corrections in the SMEFT is so highly correlated.
The invariance is manifest in a particular operator basis, but largely hidden in other formalisms. Nevertheless the invariance follows from a simple scaling argument
and its existence is input parameter scheme independent. In order to check this, we have developed a $\{\hat{m}_W, \hat{m}_Z,\hat{G}_F\}$ input parameter scheme for global SMEFT fits, and applied it to
a global analysis of $\bar{\psi} \psi \rightarrow  \bar{\psi} \psi$ and  $\bar{\psi} \psi \rightarrow  \bar{\psi} \psi \,  \bar{\psi} \psi$ scattering data,
finding some scheme dependence in the conclusions. We have also discussed why the adoption of a $\{\hat{m}_W, \hat{m}_Z,\hat{G}_F\}$ input parameter
scheme has theoretical advantages as the SMEFT is further developed.

If a formalism is used to globally fit the data in the SMEFT that makes this reparameterization invariance non manifest,
then it is essential that the correlations of the Wilson coefficients, or a power counting assumption, is not simultaneously assumed to be inconsistent with
the consequences of the reparameterization invariance in order to obtain constraints that are basis independent in the SMEFT. This is already the case in global analyses
when considering  $\bar{\psi} \psi \rightarrow  \bar{\psi} \psi$ and  $\bar{\psi} \psi \rightarrow  \bar{\psi} \psi \,  \bar{\psi} \psi$
data. Although this can be done in operator bases in a fairly direct fashion, it is
not clear how a mass eigenstate parameter formalism and corresponding fits can define such a theoretical correlation matrix\footnote{See Ref.~\cite{Berthier:2016tkq}
for discussion and an attempt to define such a correlation matrix, however, we caution that it does not seem possible to prove that using the bilinear nature of the covariance matrix
is comparable with the EOM consequences of the reparameterization invariance property.} to ensure the consequences of the 
reparameterization invariance in Wilson coefficient relationships is not explicitly broken by assumption, instead of the consistent use of the data. These challenges can be further emphasised and introduce further inconsistencies with the inclusion of the vast LHC data set that is being 
recorded and reported in EFT analyses. 

Irrespective of what approach is used, the results of this work favour the use of EFT formalisms that do not obscure the physical consequences of the relations in Eqn.~\ref{EOMrelations} in order to
obtain a consistent global constraint picture on physics beyond the SM combining LEP, low energy and LHC data.

\section*{Acknowledgements}
MT and IB acknowledge generous support from the Villum Fonden and partial support by the Danish National Research Foundation (DNRF91). 
We thank Mikkel Bj\o rn and Laure Berthier for their fundamental contribution to the global fit feeding into this analysis and particularly Mikkel Bj\o rn for providing support with the code.
MT also thanks William Shepherd for patience and insightful and helpful discussions related to this work and comments on the draft.
We thank A. Freitas for suggesting the "good enough" approach adopted for $\{\hat{m}_W, \hat{m}_Z,\hat{G}_F\}$ scheme predictions in this preliminary study, and very helpful correspondence.
 We also thank Emil Bjerrum-Bohr, Poul Damgaard,
Aneesh Manohar, Andr\'e Tinoco Mendes, Giampiero Passarino and Witold Skiba for comments on the draft and useful discussions. MT thanks
Mark Wise and Benjamin Grinstein for explanations when pursuing an (unsuccessful) alternative explanation of the flat directions.
MT thanks Giampiero Passarino 
for collaboration leading to Ref.~\cite{deFlorian:2016spz,Passarino:2016pzb}, an initial exposure to the $\{\hat{m}_W, \hat{m}_Z,\hat{G}_F\}$ scheme, and related explanations.

\appendix

\section{Jacobian relations between input parameter schemes}
The mapping of the shifts in observables in the SMEFT in the $\{\hat{\alpha}, \hat{m}_Z,\hat{G}_F\}$-scheme into the $\{\hat{m}_W, \hat{m}_Z,\hat{G}_F\}$-scheme can be directly inferred as follows. The total shift in an observable $X$ due to all operators in the SMEFT, computed in a scheme of input parameters $\{y_i\}$ we denote as 
\begin{equation}
 (\delta X) = (\delta X)_d +  \frac{\Delta X}{\Delta y_i} \Delta y_i\,.
\end{equation} 
Here we are denoting a linearized variation at leading order in the power counting of the SMEFT with the notation $\Delta$ and $\Delta y_i$ denotes the correction in an input parameter $\{y_i\}$
of this order so that $\Delta y_i = \hat{y}_i- \bar{y_i}$. $(\delta X)_d$ denotes a direct $\mathcal{L}_6$ operator contribution to an observable due to an operator in $\mathcal{L}_6$, present in any scheme.
This can be easily translated into another input parameters set $\{z_j\}$ via
 \begin{equation}
(\delta X) =  (\delta X)_d +  \frac{\Delta X}{\Delta y_i} \Delta y_i =  (\delta X)_d + \frac{\Delta X}{\Delta z_j}\frac{\Delta z_j}{\Delta y_i}\Delta y_i 
 =  (\delta X)_d + \frac{\Delta X}{\Delta z_j} (\Delta z_j)_{\sum y_i},
 \end{equation} 
 where $(\Delta z_j)_{\sum y_i}$ denotes the shift in the quantity $z_j$ computed in the $y_i$ scheme due to input parameter dependence.
 In the input parameter schemes used in this paper we take 
 \bea
 y_i=\{\bar{G}_F,\bar{m}_Z,\bar{m}_W\}\quad \text{and} \quad z_j=\{\bar{G}_F, \bar{m}_Z,\bar{\a}\}.
 \eea
The overlap of $y_i,z_j$ and the orthogonality of the input parameters leads to a shift in translating from the 
$\{\hat{\alpha}, \hat{m}_Z,\hat{G}_F\}$-scheme into the $\{\hat{m}_W, \hat{m}_Z,\hat{G}_F\}$-scheme being given by 
\bea
\frac{\Delta X}{\Delta y_i} \Delta y_i  - \frac{\Delta X}{\Delta z_j} (\Delta z_j)_{\sum y_i}.
\eea
For our case, this expression simplifies to a correction of the form 
\bea
\frac{\Delta X }{\Delta \bar{\a}}(\Delta \bar{\a})_{m_W}.
\eea
We use this simple cross check of the results reported obtained by direct calculation.
This simple relationship is somewhat accidental in the input parameter sets examined here, and follows from 
\bea\label{shiftidentities}
(\delta \bar{G}_F)_{m_W}&=&(\d \bar{G}_F)_{\a} \equiv (\Delta \bar{G}_F), 
\eea
\bea\label{shiftidentities2}
(\delta \bar{m}_Z^2)_{m_W}&=&(\d \bar{m}_Z^2)_{\a} \equiv (\Delta \bar{m}_Z^2). 
\eea

\section{\texorpdfstring{$\{\hat{m}_W, \hat{m}_Z,\hat{G}_F\}$}{\{mW,mZ,GF\}} inputs scheme in the HEFT}\label{appendix_heft}

In this Appendix we develop the $\{\hat{m}_W, \hat{m}_Z,\hat{G}_F\}$ input scheme for the HEFT Lagrangian, deriving the corresponding expressions of the core shifts parameters. We employ the basis of Ref.~\cite{Brivio:2016fzo} in the $\rm U(3)^5$ flavour symmetric limit and, unlike in the SMEFT case, we use a notation with dimensionless Wilson coefficients $c_i$, writing explicitly the suppression scale $\Lambda$ when necessary. Details about the definition of the fields and operators of the HEFT Lagrangian can be found in Ref.~\cite{Brivio:2016fzo}.

The input parameter shifts read in this case:
\bea
\d G_F &=& 
-\frac{64 \pi^2}{2 \hat{G}_F^2} \frac{r_2^\ell-r_5^\ell}{\Lambda^2}, \\
\dfrac{\d m_Z^2}{\hat{m}_Z^2} &=& 
-c_T-2\frac{\hat{m}_W}{\hat{m}_Z}\sqrt{1-\frac{\hat{m}_W^2}{\hat{m}_Z^2}} c_1, \\
\dfrac{\d m_W^2}{\hat{m}_W^2} &=& -2c_{12},
\eea
where it is worth noting that the operator $\mathcal{P}_{12}=({\rm Tr}(\mathbf{T}W_{\mu\nu}))^2\mathcal{F}_{12}(h)$, which is equivalent to the dimension-8 operator $(H^\dagger W_{\mu\nu}H)^2$, introduces a shift in the $\hat m_W$ parameter, which is identically vanishing in the SMEFT case when only including $\mathcal{L}_6$ corrections.

The shift in the Weinberg angle is consequently given by
\bea
\d s^2_{\theta} &=&-2c_T\frac{\hat{m}_W^2}{\hat{m}_Z^2}-2c_1\frac{\hat{m}_W}{\hat{m}_Z}\sqrt{1-\frac{\hat{m}_W^2}{\hat{m}_Z^2}}+4c_{12}\frac{\hat{m}_W^2}{\hat{m}_Z^2}.
\eea

The shifts in the $Z$ couplings to fermions can be expressed in the notation of Eq.~\ref{Zcouplngs.shift}, where, for the HEFT theory:
\begin{equation}
 \d \bar{g}_Z = -\frac{1}{\sqrt2} \, \d G_F-\frac{1}{2}\frac{\d m_Z^2}{\hat{m}_Z^2}-s_{2\hat\theta}c_{1}
 =c_T+\frac{16\pi^2}{\sqrt2 \hat{G}_F}\frac{r_2^\ell-r_5^\ell}{\Lambda^2}.
\end{equation}
As in the SMEFT case, the universal shift $\delta\bar g_Z$ is unchanged when moving from the $\{\alpha, \hat{m}_Z,\hat{G}_F\}$ to the $\{\hat{m}_W, \hat{m}_Z,\hat{G}_F\}$ scheme.
The direct contributions $\Delta_{V,A}^f$ read
\begin{align}
 (\Delta^\ell_V)_{pr} &= \begin{cases}\frac{1}{2}(-n^\ell_{\mathbf{V}} +2n^\ell_{\{\mathbf{V}\mathbf{T}\}}-n^\ell_{\mathbf{T}\mathbf{V}\mathbf{T}}+2n_2^\ell)_{pr}& (p\neq r)\\
		       (n^\ell_2)_{rr}& (p=r)\end{cases},\\
 (\Delta^\ell_A)_{pr} &= \begin{cases}\frac{1}{2}(-n^\ell_{\mathbf{V}} +2n^\ell_{\{\mathbf{V}\mathbf{T}\}}-n^\ell_{\mathbf{T}\mathbf{V}\mathbf{T}}-2n_2^\ell)_{pr}& (p\neq r)\\
		       -(n^\ell_2)_{rr}& (p=r)\end{cases},\\
 (\Delta^\nu_V)_{pr} &= (\Delta^\nu_A)_{pr} =\begin{cases}\frac{1}{2}(n^\ell_{\mathbf{V}} +2n^\ell_{\{\mathbf{V}\mathbf{T}\}}+n^\ell_{\mathbf{T}\mathbf{V}\mathbf{T}})_{pr}& (p\neq r)\\
                                              0& (p=r)
                                             \end{cases},\\
 (\Delta^u_V)_{pr} &= \frac{1}{2}\left(n^Q_1+n^Q_2+2n^Q_5+2n^Q_6+n^Q_7+n^Q_8\right)_{pr},\\
 (\Delta^u_A)_{pr} &= \frac{1}{2}\left(n^Q_1-n^Q_2+2n^Q_5-2n^Q_6+n^Q_7-n^Q_8\right)_{pr},\\
 (\Delta^d_V)_{pr} &= \frac{1}{2}\left(-n^Q_1-n^Q_2+2n^Q_5+2n^Q_6-n^Q_7-n^Q_8\right)_{pr},\\
 (\Delta^d_A)_{pr} &= \frac{1}{2}\left(-n^Q_1+n^Q_2+2n^Q_5-2n^Q_6-n^Q_7+n^Q_8\right)_{pr},
\end{align}
where $p,r$ are flavor indices and we have denoted by $n^\ell_{\mathbf{V}},\,n^\ell_{\{\mathbf{V}\mathbf{T}\}},\,n^\ell_{\mathbf{T}\mathbf{V}\mathbf{T}}$ the Wilson coefficients, respectively, of the operators
$$
i\bar{L}_{L,p}\g_\mu\mathbf{V}^\mu L_{L,r},\quad
i\bar{L}_{L,p}\g_\mu\{\mathbf{V}^\mu,\mathbf{T}\} L_{L,r},\quad
i\bar{L}_{L,p}\g_\mu\mathbf{T}\mathbf{V}^\mu\mathbf{T} L_{L,r}.
$$
The flavor diagonal components of these structures, that correspond to $\mathcal{Q}_{Hl}^{(1)}$ and $\mathcal{Q}_{Hl}^{(3)}$ in the SMEFT, were removed from the basis of Ref.~\cite{Brivio:2016fzo} and traded for bosonic operators. This explains why the shifts in the flavor diagonal $Z$-lepton couplings are much simplified compared to the $Z$-quark couplings.

The shifts in the $W^\pm$ couplings, in the normalization of Eq.~\ref{Wcouplings.def} are
\begin{align} 
 \d (g_V^{W_\pm,\ell})_{pr}= \d (g_A^{W_\pm,\ell})_{pr} &= \begin{cases} \left(n^\ell_{\mathbf{V}}-n^\ell_{\mathbf{T}\mathbf{V}\mathbf{T}}+2i n^\ell_1\right)_{pr}- \frac{ \delta G_F}{2\sqrt{2}}& (p\neq r) \\
                           (2i n^\ell_1)_{rr}- \frac{ \delta G_F}{2\sqrt{2}}& (p= r)                                  
                                                           \end{cases},\\
 \d (g_V^{W_\pm,q})_{pr}&= \left(n^Q_1-n^Q_7 +n^Q_2-n^Q_8+ 2 i (n^Q_3+n^Q_4)\right)_{pr} - \frac{ \delta G_F}{2\sqrt{2}},\\
 \d (g_A^{W_\pm,q})_{pr} &= \left(n^Q_1-n^Q_7 -n^Q_2+n^Q_8 + 2 i (n^Q_3-n^Q_4)\right)_{pr} - \frac{ \delta G_F}{2\sqrt{2}}.
\end{align} 
Note that in the HEFT formalism it is possible to have $W^\pm$ couplings to righthanded quark currents at the first order in the power counting: these are parameterized by the coefficients $n^Q_2$ and $n^Q_8$. The same is not true in the lepton sector due to the absence of righthanded neutrinos. Finally, the coefficients $n^\ell_1$, $n^Q_3$, $n^Q_4$ are intrinsically CP odd.

The effective photon couplings are proportional to $\hat e(1+\delta e/\hat e)$, where
\bea
\frac{\delta e}{\hat{e}} \equiv \frac{\delta \alpha}{2 \, \hat{\alpha}} =
-\frac{\delta G_F}{\sqrt{2}} + \dfrac{\d m_Z^2}{\hat{m}_Z^2} \frac{\hat{m}_W^2}{2 \, (\hat{m}_W^2 - \hat{m}_Z^2)}
-\frac{\delta m_W^2}{\hat{m}_W^2}\frac{2\hat{m}_W^2-\hat{m}_Z^2}{2(\hat{m}_W^2-\hat{m}_Z^2)}
+2 c_1 \frac{\hat{m}_W}{\hat{m}_Z} \sqrt{1-\frac{\hat{m}_W^2}{\hat{m}_Z^2}}.\qquad
\eea

Finally, variations in the triple gauge boson interaction can be expressed in the parameterization of Ref.~\cite{Hagiwara:1986vm} as
\begin{equation}
\begin{aligned}
\frac{\LL_{WWV,eff}}{-i \, \hat{g}_{WWV}} =& 
g_1^V \Big( W^+_{\mu\nu} W^{- \mu} V^{\nu} - W^+_{\mu} V_{\nu} W^{- \mu\nu} \Big) + \kappa_V W_\mu^+ W_\nu^- V^{\mu\nu} +\frac{i\lambda_V}{\hat{m}_W^2}V^{\mu\nu}W^{+\rho}_\nu W^-_{\rho\mu}\\
& - ig_5^V \epsilon^{\mu\nu\rho\sigma} \left(W_\mu^+\de_\rho W^-_\nu-W_\nu^-\de_\rho W^+_\mu\right)V_\s ,
\end{aligned} 
\end{equation}
and their expression in terms of the HEFT coefficients are the following:
\begin{align}
 \d g_1^\gamma &= 2c_1\frac{\hat{m}_W}{\sqrt{\hat{m}_Z^2-\hat{m}_W^2}}+2c_{12}\frac{2 \hat{m}_W^2-\hat{m}_Z^2}{\hat{m}_W^2-\hat{m}_Z^2}-c_T\frac{\hat{m}_W^2}{\hat{m}_W^2-\hat{m}_Z^2}+\frac{8 \sqrt{2} \pi ^2 }{\hat{G}_F\Lambda^2}(r_2^\ell-r_5^\ell), \\
 \d g_1^Z &=4c_{12}-2c_1\frac{\hat{m}_Z}{\hat{m}_W}\sqrt{1-\frac{\hat{m}_W^2}{\hat{m}_Z^2}}-c_T+\frac{8 \sqrt{2} \pi ^2}{\hat{G}_F\Lambda^2}(r_2^\ell-r_5^\ell)+\frac{\sqrt{\hat{G}_F}}{2^{3/4}\pi}\frac{\hat{m}_Z^2}{\hat{m}_W}c_{13},\\
 \d \kappa_\gamma &= -2c_{12}\frac{\hat{m}_Z^2}{\hat{m}_W^2-\hat{m}_Z^2}-c_T\frac{\hat{m}_W^2}{\hat{m}_W^2-\hat{m}_Z^2}+\frac{8 \sqrt{2} \pi ^2 }{\hat{G}_F\Lambda^2}(r_2^\ell-r_5^\ell)+\frac{\sqrt{\hat{G}_F}\hat{m}_W}{2^{3/4}\pi}\left(\frac{2c_2}{t_{\hat\theta}}+c_3+2c_{13}\right)
,\\
 \d \kappa_Z &= -c_T+\frac{8 \sqrt{2} \pi ^2}{\hat{G}_F\Lambda^2}(r_2^\ell-r_5^\ell)+\frac{\sqrt{\hat{G}_F}\hat{m}_W}{2^{3/4}\pi}\left(-2t_{\hat\theta}c_2+c_3+2c_{13}\right),\\
 \d \lambda_\gamma &=6 \, s_{\hat{\theta}} \, \frac{\hat{m}^2_W}{\hat{g}_{WWA}} \,  c_{WWW},\\
 \d \lambda_Z &=6 \, c_{\hat{\theta}} \, \frac{\hat{m}^2_W}{\hat{g}_{WWZ}} \,  c_{WWW},\\
 \d g_5^\gamma &=0,\\
 \d g_5^Z &=\frac{\sqrt{\hat{G}_F}}{2^{3/4}\pi}\frac{\hat{m}_Z^2}{\hat{m}_W}c_{14}.
\end{align}
Compared to the shifts obtained in the SMEFT, more independent HEFT operators contribute to TGCs. This is partly due to a different basis choice for effects equivalent to dimension-6 invariants: as an example, the HEFT basis of Ref.~\cite{Brivio:2016fzo} contains the operators ${\mathcal{P}_2\sim B_{\mu\nu}{\rm Tr}(\mathbf{T}[\mathbf{V}^\mu,\mathbf{V}^\nu])\mathcal{F}_2(h)}$ and ${\mathcal{P}_3\sim {\rm Tr}(W_{\mu\nu}[\mathbf{V}^\mu,\mathbf{V}^\nu])\mathcal{F}_3(h)}$, whose linear ``siblings'' are the structures $D^\mu H^\dagger B_{\mu\nu} D^\nu H$ and $D^\mu H^\dagger W_{\mu\nu} D^\nu H$ respectively, that were not retained in the SMEFT basis of Ref.~\cite{Grzadkowski:2010es}.
In addition, triple gauge couplings receive the contribution of HEFT operators that correspond to terms of dimension $d\geq 8$ in the SMEFT formalism, such as $\mathcal{P}_{13}\sim{\rm Tr}(\mathbf{T}W_{\mu\nu}){\rm Tr}(\mathbf{T}[\mathbf{V}^\mu,\mathbf{V}^\nu])\mathcal{F}_{13}(h)$ and $\mathcal{P}_{14}\sim\varepsilon^{\mu\nu\rho\lambda}{\rm Tr}(\mathbf{T}\mathbf{V}_{\mu}){\rm Tr}(\mathbf{V}_\nu W_{\rho\lambda})\mathcal{F}_{14}(h)$. In particular, the latter gives a non-vanishing $\delta g_5^Z$.

Finally, it is worth noting that due to the contribution of the operator $\mathcal{P}_{12}$, the SMEFT relationship ${\d \kappa_Z = \d g_1^Z- t_{\theta}^2 (\d \kappa_\gamma - \d g_1^\gamma)}$  does not hold for the HEFT Lagrangian even at leading order.
\section{Numerical global fit results}\label{appendixresults}
\begin{table}[h!]
\renewcommand{\arraystretch}{1.3}
 \begin{tabular}{|l*4{|r@{ $\pm$ }l|r@{ $\pm$ }l}|}
  \toprule
  \multirow{2}{*}{$C_i\times\frac{\bar{v}_T^2}{\Lambda^2}$}& \multicolumn{4}{c}{$\{\hat{\alpha},\hat{m}_Z, \hat{G}_F\}$ scheme} & \multicolumn{4}{|c|}{$\{\hat{m}_W,\hat{m}_Z, \hat{G}_F\}$ scheme}\\
  & \multicolumn{2}{c}{(0\%)}& \multicolumn{2}{c}{(1\%)} & \multicolumn{2}{|c}{(0\%)}& \multicolumn{2}{c|}{(1\%)}\\\midrule
$C_{He}$	&47.	&25.    &34.	&32.  	&36.	&21.	&26.	&27.\\
$C_{Hu}$	&-31.	&17.	&-22.	&22.	&-23.	&14.	&-16.	&18.\\
$C_{Hd}$	&12.8	&8.4   	&8.	&11.	&8.3	&6.9	&4.9	&9.2\\
$C_{Hl}^{(1)}$	&24.	&13. 	&17.	&16.	&18.	&10.	&13.	&13.\\
$C_{Hl}^{(3)}$	&81.	&47.   	&71.	&50.	&68.	&42.	&61.	&44.\\
$C_{Hq}^{(1)}$	&-7.8	&4.2   	&-5.7	&5.4	&-6.0	&3.5	&-4.5	&4.5\\
$C_{Hq}^{(3)}$	&80.	&47.  	&71.	&50.	&67.	&42.	&61.	&44.\\
$C_{HWB}$	&3.4	&6.5  	&-5.	&13.	&-2.3	&7.7	&-8.	&12.\\
$C_{HD}$	&-94.	&51. 	&-67.	&65.	&-72.	&41.	&-52.	&54.\\
$C_{ll}$	&-0.19	&0.18	&-0.7	&1.0	&-0.42	&0.56	&-0.8	&1.1\\
$C_{ee}$	&9.1	&6.3 	&8.6	&7.4	&5.3	&9.0	&6.7	&9.4\\
$C_{le}$	&4.4	&5.5 	&4.6	&5.6	&3.6	&5.5	&3.9	&5.8\\
$C_{W}$		&120.	&72. 	&110.	&75.	&99.	&62.	&93.	&65.\\
  \bottomrule
 \end{tabular}
\caption{Best fit values and corresponding $1\sigma$ confidence regions for $\Delta_{\rm SMEFT}=\{0\%, 1\%\}$ and for the two input parameter schemes considered in this work. The numbers have been obtaining after profiling the $\chi^2$ over the other parameters and they have been multiplied by a factor 100. }\label{tab.fitresults_numeric_profiling}
\end{table}

\begin{table}[h!]
\renewcommand{\arraystretch}{1.3}
 \begin{tabular}{|l*4{|r@{ $\pm$ }l|r@{ $\pm$ }l}|}
  \toprule
  \multirow{2}{*}{$C_i\times\frac{\bar{v}_T^2}{\Lambda^2}$}& \multicolumn{4}{c}{$\{\hat{\alpha},\hat{m}_Z, \hat{G}_F\}$ scheme} & \multicolumn{4}{|c|}{$\{\hat{m}_W,\hat{m}_Z, \hat{G}_F\}$ scheme}\\
  & \multicolumn{2}{c}{(0\%)}& \multicolumn{2}{c}{(1\%)} & \multicolumn{2}{|c}{(0\%)}& \multicolumn{2}{c|}{(1\%)}\\\midrule
$C_{He}$	&-0.047	&0.036	&-0.064	&0.079	&-0.054	&0.037	&-0.104	&0.092\\		
$C_{Hu}$	&0.06	&0.25	&0.45	&0.87	&-0.06	&0.25	&0.462	&1.036\\		
$C_{Hd}$	&-0.35	&0.33	&-2.1	&1.1	&-0.152	&0.33	&-2.4	&1.3  \\
$C_{Hl}^{(1)}$	&0.016	&0.025	&-0.07	&0.10	&0.018	&0.026	&-0.109	&0.11 \\
$C_{Hl}^{(3)}$	&-0.013	&0.025	&0.019	&0.054	&-0.009	&0.039	&-0.12	&0.11 \\
$C_{Hq}^{(1)}$	&0.05	&0.10	&0.05	&0.41	&0.01	&0.11	&0.05	&0.42	\\	
$C_{Hq}^{(3)}$	&0.013	&0.037	&0.21	&0.29	&-0.005	&0.039	&0.21	&0.30	\\
$C_{HWB}$	&-0.008	&0.020	&0.015	&0.029	&-0.046	&0.053	&-0.050	&0.061\\	
$C_{HD}$	&-0.058	&0.051	&0.01	&0.11	&-0.075	&0.059	&-0.066	&0.066\\		
$C_{ll}$	&0.019	&0.044	&-0.053	&0.074	&0.011	&0.094	&-0.79	&0.58	\\	
$C_{ee}$	&12.4	&4.6	&12.0	&5.4	&11.9	&4.4	&11.5	&5.2\\
$C_{le}$	&9.8	&4.0	&8.8	&4.2	&9.4	&3.9	&8.5	&4.0\\
$C_{W}$		&1.8	&4.5	&1.9	&4.5    &1.9	&4.4    &2.0	&4.5 \\

  \bottomrule
 \end{tabular}
\caption{Best fit values and corresponding $1\sigma$ confidence regions for $\Delta_{\rm SMEFT}=\{0\%, 1\%\}$ and for the two input parameter schemes considered in this work. These numbers have been obtained minimizing the $\chi^2$ with one parameter at a time (despite the non-minimal character of the SMEFT \cite{Jiang:2016czg}), and they have been multiplied by a factor 100.}\label{tab.fitresults_numeric_oneatatime}
\end{table}

\clearpage

\section{Addendum: independent treatment of \texorpdfstring{$U(3)^5$}{U(3)\^{ }5} contractions for \texorpdfstring{$\mathcal{Q}_{ll}$}{Qll}}

The main text adopts an overly restrictive form of a $\rm U(3)^5$ limit by not allowing
two independent flavor contractions admitted by the operator $\Q_{ll}$
in the $\rm U(3)^5$ flavor symmetric limit~\cite{Cirigliano:2009wk}. Defining
\begin{equation*}
\mathcal{L}_{\rm SMEFT}\supset\left[C_{ll}\,\delta_{mn}\delta_{op}+C_{ll}'\,\delta_{mp}\delta_{no}\right](\bar l_m\gamma_\mu l_n)(\bar l_o\gamma^\mu l_p) ,
\end{equation*}
both $C_{ll}$ and $C_{ll}'$ are allowed to be independent parameters in the $\rm U(3)^5$  flavour symmetric limit.
The discussion above uses the same parameter $C_{ll}$ in both terms, which is overly restrictive.
This leads to $C_{ll} \rightarrow C_{ll}'$ in the equations~(3.4),~(3.15), (3.22)-(2.25) that read respectively:
\begin{align}
 \d G_F &=
\dfrac{1}{\sqrt{2} \, \hat{G}_F}\left(\sqrt2 C_{H\ell}^{(3)}-\dfrac{1}{\sqrt2}C_{ll}'\right),\\
 \d \bar{g}_Z &= -\frac{1}{\sqrt2} \, \d G_F-\frac{1}{2}\frac{\d m_Z^2}{\hat{m}_Z^2}+\frac{s_{\hat\theta} c_{\hat\theta}}{\sqrt2 \hat{G}_F}C_{HWB}
 =-\frac{1}{4\sqrt2 \hat{G}_F}\left( C_{HD}+4 \, C_{H\ell}^{(3)}-2 C_{ll}'\right),\\
 \d g_1^\gamma &= \frac{1}{4\sqrt2 \hat{G}_F}\left(C_{HD}\frac{\hat{m}_W^2}{\hat{m}_W^2-\hat{m}_Z^2}-4C_{H\ell}^{(3)}+2C_{ll}'-C_{HWB}\frac{4\hat{m}_W}{\sqrt{\hat{m}_Z^2-\hat{m}_W^2}}\right), \\
 \d g_1^Z &=\frac{1}{4\sqrt2 \hat{G}_F}\left(C_{HD}-4C_{H\ell}^{(3)}+2 C_{ll}'+4\frac{\hat{m}_Z}{\hat{m}_W}\sqrt{1-\frac{\hat{m}_W^2}{\hat{m}_Z^2}}C_{HWB}\right),\\
 \d \kappa_\gamma &=  \frac{1}{4\sqrt2 \hat{G}_F}\left(C_{HD}\frac{\hat{m}_W^2}{\hat{m}_W^2-\hat{m}_Z^2}-4C_{H\ell}^{(3)}+2C_{ll}'\right),\\
 \d \kappa_Z &= \frac{1}{4\sqrt2 \hat{G}_F}\left(C_{HD}-4C_{H\ell}^{(3)}+2C_{ll}' \right),
\end{align}
The list in Eq.~(3.37) should also include $C_{ll}'$:
\begin{equation}
\tilde{C}_i \equiv \frac{\bar{v}_T^2}{\Lambda^2} \{
 C_{He}, C_{Hu}, C_{Hd}, \CHls, \CHlt, \CHqs, \CHqt, C_{HWB}, C_{HD}, C_{ll}, C_{ll}', C_{ee}, C_{le}
 \},
\end{equation}
and the number of Wilson coefficients in the text after Eq.~(3.45) is then 21.

\begin{figure}[t]
 \centering
 \includegraphics[height=12cm, trim ={0 0 4.6cm 0}, clip]{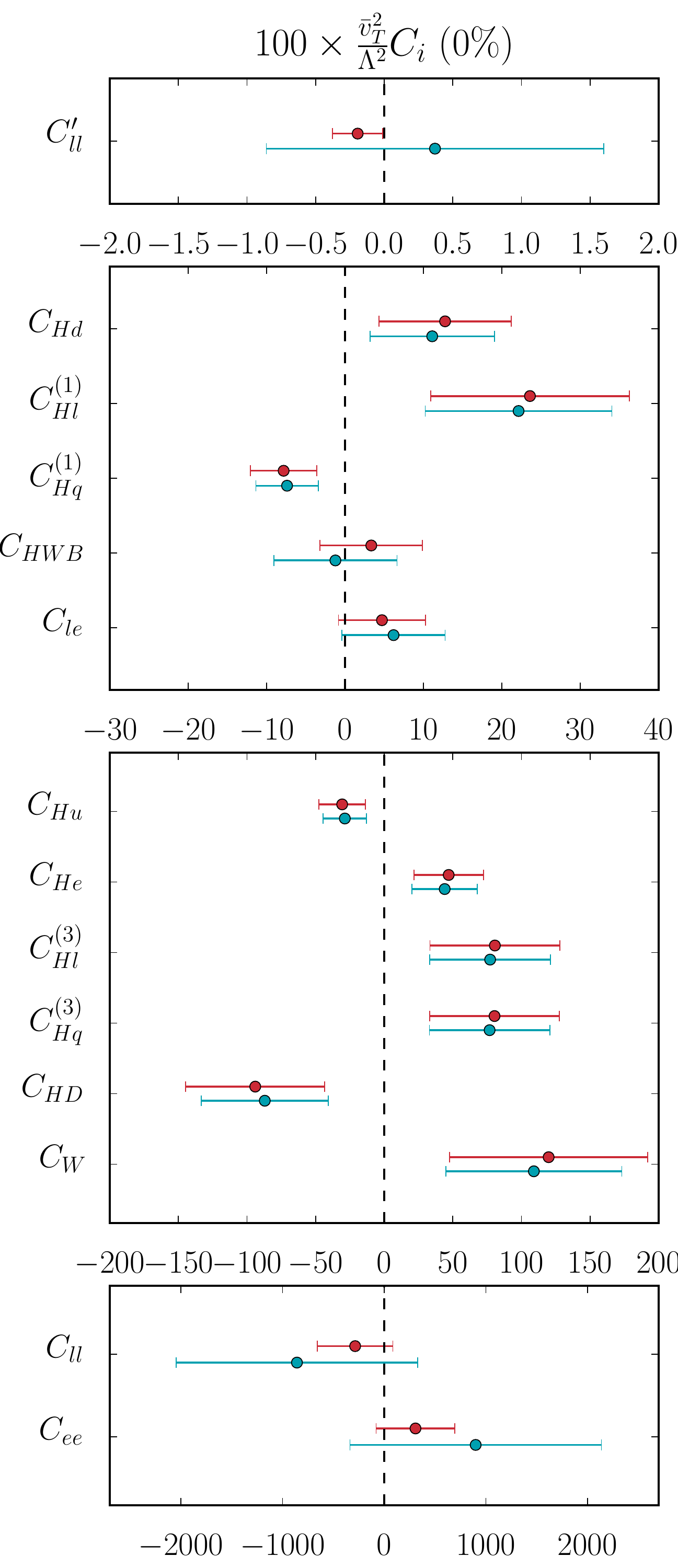}
 \includegraphics[height=12cm]{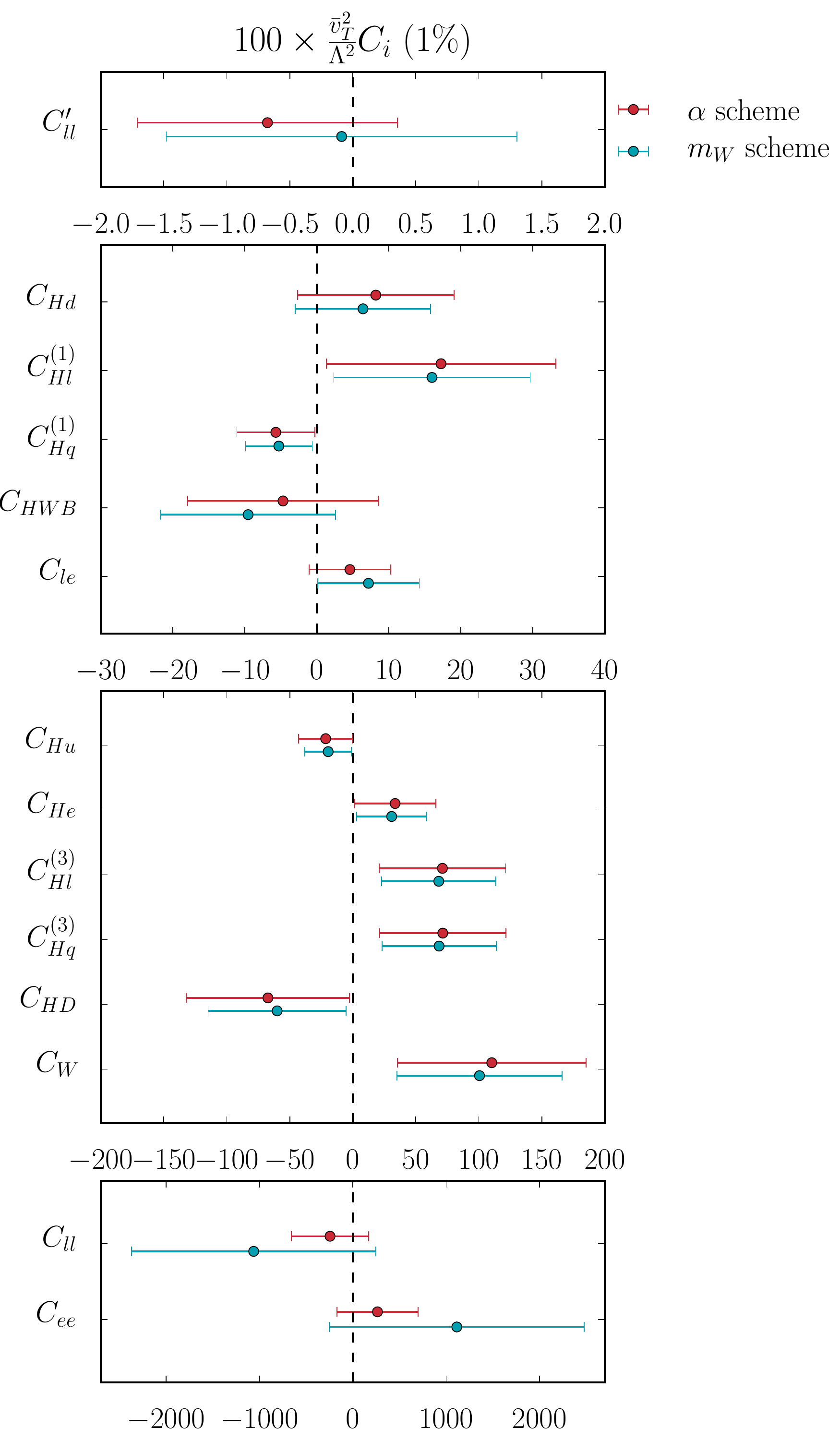}
 \caption{Best fit values of the Wilson coefficients (scaled by a factor 100) and corresponding $\pm1\sigma$ confidence regions obtained after profiling away the other parameters. Red (blue) points were obtained in the $\{\hat\alpha \,(\hat{m}_W), \hat{m}_Z,\hat{G}_F\}$ input parameter scheme. The plot to the left has been obtained assuming $\Delta_{\rm SMEFT}=0$, while the one to the right includes a theoretical error $\Delta_{\rm SMEFT}=0.01$. }
 \label{Fig:limits_profiling_cll1}
\end{figure}

\begin{figure}[t]
 \centering
 \includegraphics[height=12cm, trim ={0 0 4.71cm 0}, clip]{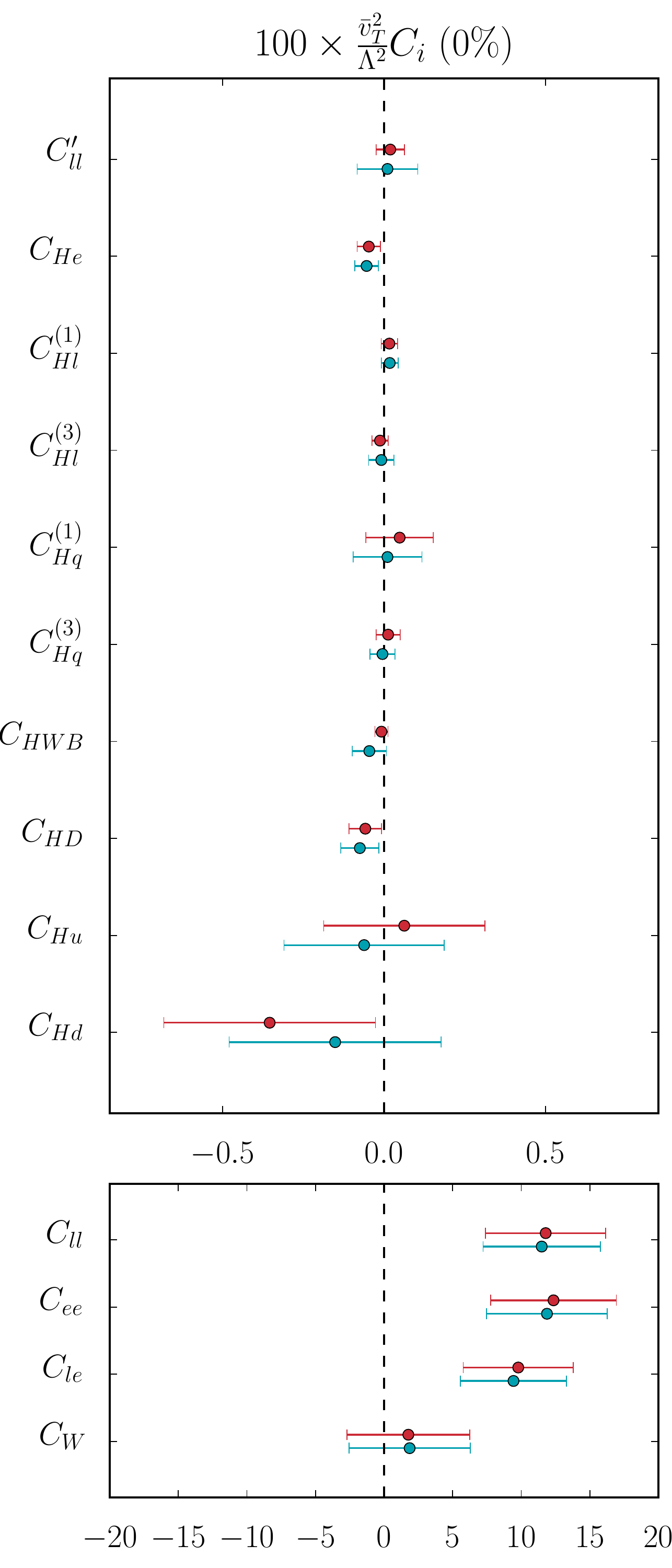}
 \includegraphics[height=12cm]{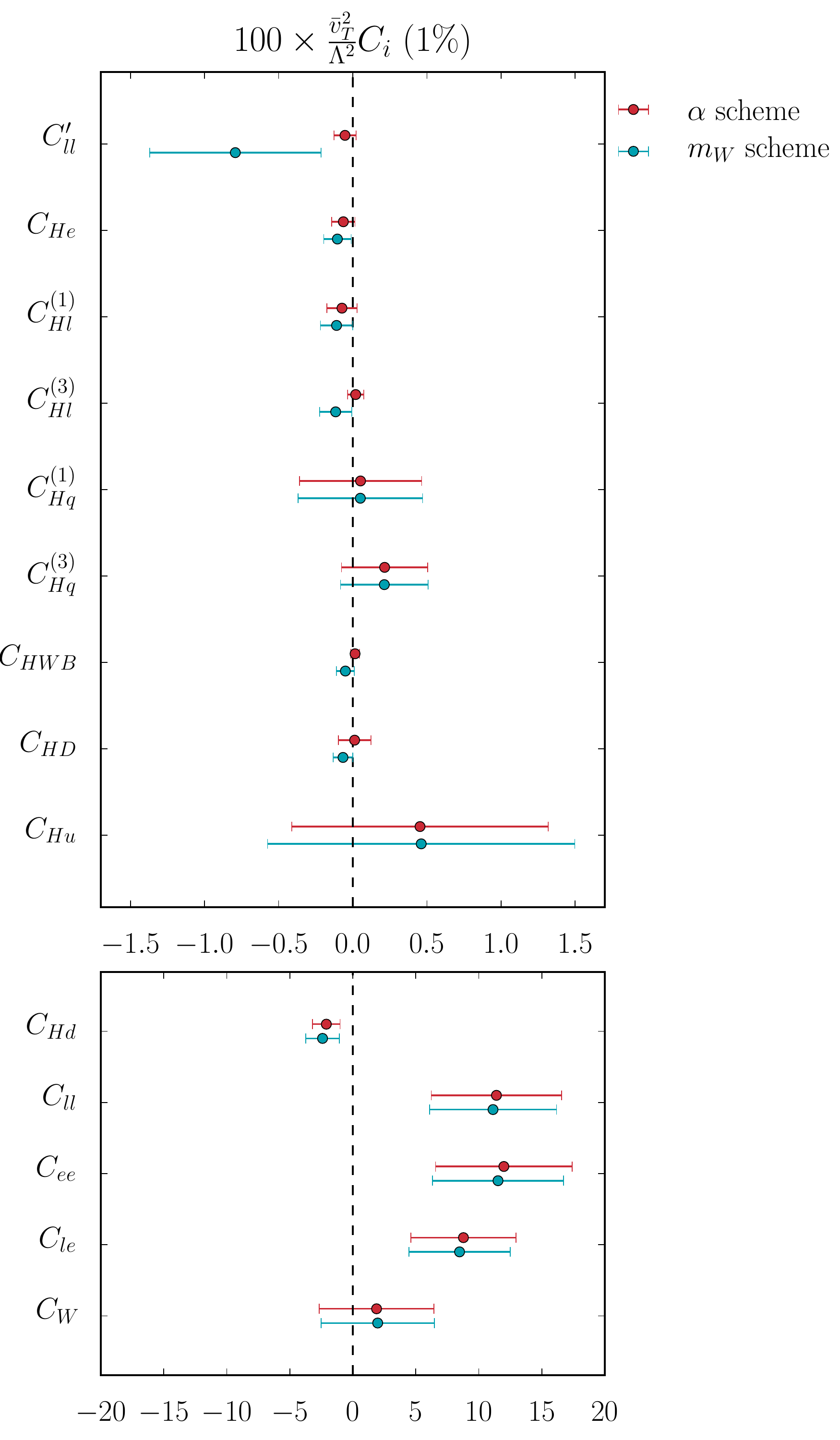}
 \caption{Best fit values of the Wilson coefficients (scaled by a factor 100) and corresponding $\pm1\sigma$ confidence regions obtained minimizing the $\Delta\chi^2$ with one parameter at a time. Red (blue) points were obtained in the $\{\hat\alpha \, (\hat{m}_W), \hat{m}_Z,\hat{G}_F\}$ input parameter scheme. The plot to the left has been obtained assuming $\Delta_{\rm SMEFT}=0$, while the one to the right includes a theoretical error $\Delta_{\rm SMEFT}=0.01$. Note that in the right plot the $x$ axis has been scaled by a factor 2 and the coefficient $C_{Hd}$ has been moved to the lower panel: increasing the theoretical error enhances the pull of the $\mathcal{A}_{FB}^{0,b}$ anomaly compared to $Z$ width data, and this relaxes by one order of magnitude the bound on this parameter.}
 \label{Fig:limits_oneatatime_cll1}
\end{figure}

\begin{figure}[t]
 \centering
 \includegraphics[width=\textwidth]{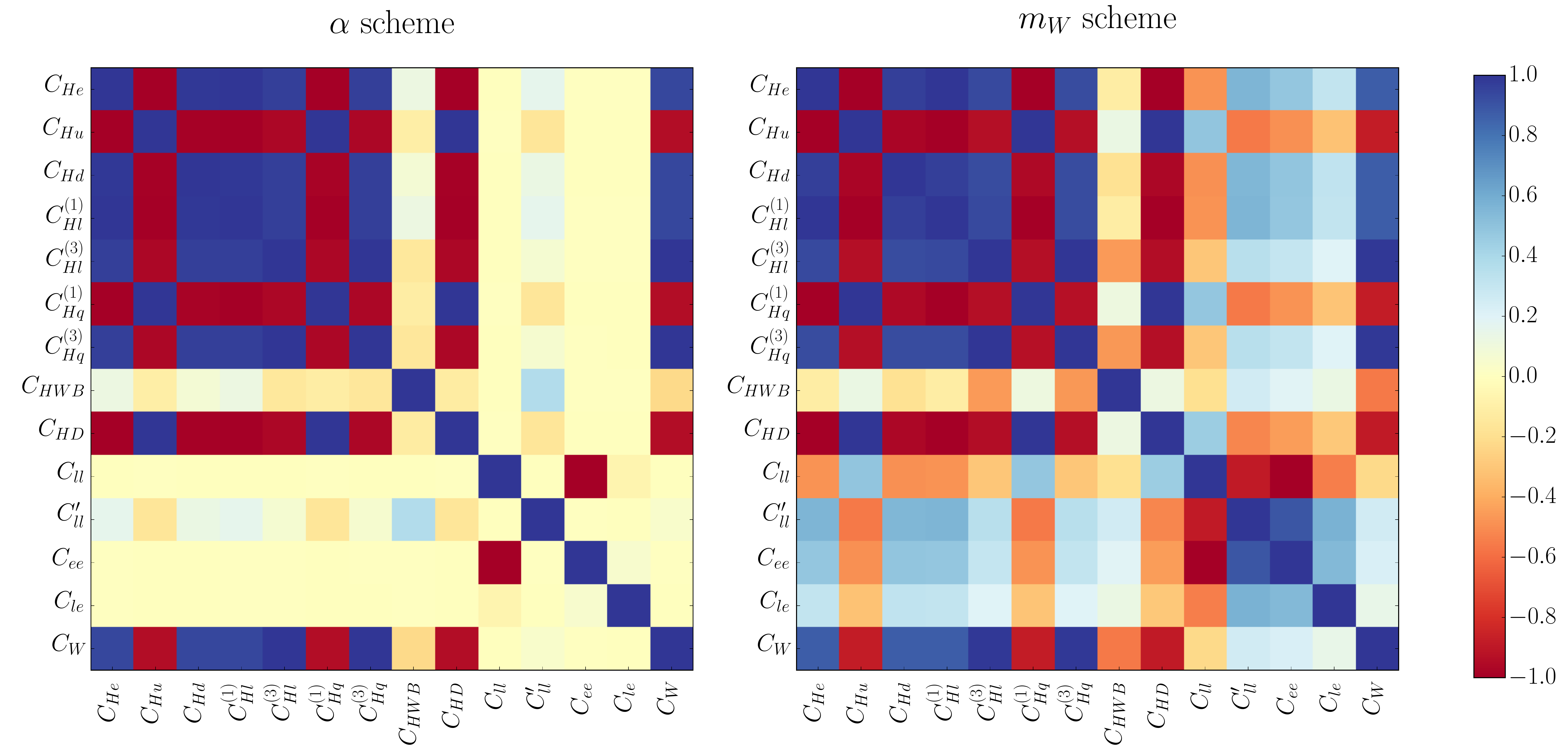}
 \caption{Color map of the correlation matrix among the Wilson coefficients, obtained assuming zero SMEFT error, for the $\{\hat{\alpha}, \hat{m}_Z,\hat{G}_F\}$ input scheme (left) and for the $\{\hat{m}_W, \hat{m}_Z,\hat{G}_F\}$ input scheme (right).}
 \label{Fig:correlation_matrices_cll1}
\end{figure}

The fit results in this case are shown in Figures~\ref{Fig:limits_profiling_cll1},~\ref{Fig:limits_oneatatime_cll1},~\ref{Fig:correlation_matrices_cll1}
and Tables~\ref{tab.fitresults_numeric_profiling_cll1},~\ref{tab.fitresults_numeric_oneatatime_cll1}.
The limits obtained minimizing the coefficients one-at-a-time are largely unchanged, while the fit results that marginalize over
the larger set of parameters are modified. A significant scheme dependence is found for $C_{ll}'$ in this case.
This coefficient enters the considered observables via shift parameters. In the $\{\hat\alpha,\hat m_Z, \hat G_F\}$-scheme
it impacts most LEPI data, and in particular $\hat m_W$.
In the $\{\hat m_W,\hat m_Z, \hat G_F\}$-scheme it affects dominantly bhabha scattering via $\delta\a$, that is less constraining.
$C_{ll}$ and $C_{ee}$ are poorly constrained and strongly anti-correlated as they both contribute to bhabha scattering only,
where they enter in a linear combination of the form\footnote{Here $c_\theta$ is the cosine of the angle between the incoming $e^-$ and the outgoing $e^+$ in bhabha scattering.} $[C_{ee}+(1+\Delta(s,c_\theta)) C_{ll}]$ where $0<\Delta(s,c_\theta)< 0.1$ at the LEP2 c.m.s. energy.
The direction $C_{ll}-C_{ee}$ is nearly unconstrained and this degeneracy is weakly broken by the kinematic dependence.
The correlations are larger in the $\{\hat{m}_W, \hat{m}_Z,\hat{G}_F\}$ scheme for the observables considered.
$C_{ll}'$ is more correlated with $C_{ll},\,C_{ee},\,C_{le}$ as bhabha scattering
provides the dominant constraint on $C_{ll}'$ in this scheme increasing correlations.
In the $\{\hat{\alpha}_W, \hat{m}_Z,\hat{G}_F\}$ scheme, $C_{ll}'$ is primarily bounded by the $m_W$ measurement,
and this allows the parameters to split in less correlated blocks, one constrained by {LEPI + WW}
production data and one by bhabha scattering.


\newpage
\begin{table}[h!]
\renewcommand{\arraystretch}{1.3}
 \begin{tabular}{|l*4{|r@{ $\pm$ }l|r@{ $\pm$ }l}|}
  \toprule
  \multirow{2}{*}{$C_i\times\frac{\bar{v}_T^2}{\Lambda^2}$}& \multicolumn{4}{c}{$\{\hat{\alpha},\hat{m}_Z, \hat{G}_F\}$ scheme} & \multicolumn{4}{|c|}{$\{\hat{m}_W,\hat{m}_Z, \hat{G}_F\}$ scheme}\\
  & \multicolumn{2}{c}{(0\%)}& \multicolumn{2}{c}{(1\%)} & \multicolumn{2}{|c}{(0\%)}& \multicolumn{2}{c|}{(1\%)}\\\midrule
$C_{He}$	&47.	&25.    &34.	&32.  	&44.	&24.	&31.	&28.\\
$C_{Hu}$	&-31.	&17.	&-22.	&22.	&-29.	&16.	&-20.	&18.\\
$C_{Hd}$	&12.8	&8.4   	&8.	&11.	&11.	&7.9	&6.4	&9.4\\
$C_{Hl}^{(1)}$	&24.	&13. 	&17.	&16.	&22.	&12.	&16.	&14.\\
$C_{Hl}^{(3)}$	&81.	&47.   	&71.	&50.	&77.	&44.	&68.	&45.\\
$C_{Hq}^{(1)}$	&-7.8	&4.2   	&-5.7	&5.4	&-7.4	&4.0	&-5.2	&4.6\\
$C_{Hq}^{(3)}$	&80.	&47.  	&71.	&50.	&77.    &44.	&69.	&45.\\
$C_{HWB}$	&3.4	&6.5  	&-5.	&13.	&-1.2	&7.9	&-10.	&12.\\
$C_{HD}$	&-94.	&51. 	&-67.	&65.	&-87.	&46.	&-60.	&55.\\

$C_{ll}$	&-286.	&371.	&-244.	&414.	&-859.	&1190.	&-1062.	&1310.\\

$C_{ll}'$	&-0.19	&0.18	&-0.7	&1.0	&-0.37	&1.2	&-0.08	&1.4\\
$C_{ee}$	&308.	&388. 	&264.	&434.	&890.	&1240.	&1114.	&1366.\\
$C_{le}$	&4.7	&5.5 	&4.6	&5.6	&6.2	&6.6	&7.1	&7.1\\
$C_{W}$		&120.	&72. 	&110.	&75.	&109.	&64.	&101.	&65.\\
  \bottomrule
 \end{tabular}
\caption{Best fit values and corresponding $1\sigma$ confidence regions for $\Delta_{\rm SMEFT}=\{0\%, 1\%\}$ and for the two input parameter schemes considered in this work. The numbers have been obtaining after profiling the $\chi^2$ over the other parameters and they have been multiplied by a factor 100. }\label{tab.fitresults_numeric_profiling_cll1}
\end{table}

\begin{table}[h!]
\renewcommand{\arraystretch}{1.3}
 \begin{tabular}{|l*4{|r@{ $\pm$ }l|r@{ $\pm$ }l}|}
  \toprule
  \multirow{2}{*}{$C_i\times\frac{\bar{v}_T^2}{\Lambda^2}$}& \multicolumn{4}{c}{$\{\hat{\alpha},\hat{m}_Z, \hat{G}_F\}$ scheme} & \multicolumn{4}{|c|}{$\{\hat{m}_W,\hat{m}_Z, \hat{G}_F\}$ scheme}\\
  & \multicolumn{2}{c}{(0\%)}& \multicolumn{2}{c}{(1\%)} & \multicolumn{2}{|c}{(0\%)}& \multicolumn{2}{c|}{(1\%)}\\\midrule
$C_{He}$	&-0.047	&0.036	&-0.064	&0.079	&-0.054	&0.037	&-0.104	&0.092\\
$C_{Hu}$	&0.06	&0.25	&0.45	&0.87	&-0.06	&0.25	&0.462	&1.036\\
$C_{Hd}$	&-0.35	&0.33	&-2.1	&1.1	&-0.152	&0.33	&-2.4	&1.3  \\
$C_{Hl}^{(1)}$	&0.016	&0.025	&-0.07	&0.10	&0.018	&0.026	&-0.109	&0.11 \\
$C_{Hl}^{(3)}$	&-0.013	&0.025	&0.019	&0.054	&-0.009	&0.039	&-0.12	&0.11 \\
$C_{Hq}^{(1)}$	&0.05	&0.10	&0.05	&0.41	&0.01	&0.11	&0.05	&0.42	\\
$C_{Hq}^{(3)}$	&0.013	&0.037	&0.21	&0.29	&-0.005	&0.039	&0.21	&0.30	\\
$C_{HWB}$	&-0.008	&0.020	&0.015	&0.029	&-0.046	&0.053	&-0.050	&0.061\\
$C_{HD}$	&-0.058	&0.051	&0.01	&0.11	&-0.075	&0.059	&-0.066	&0.066\\
$C_{ll}$	&11.8	&4.4	&11.4	&5.2	&11.9	&4.4	&11.1	&5.0	\\
$C_{ll}'$	&0.019	&0.044	&-0.053	&0.074	&0.011	&0.094	&-0.79	&0.58	\\
$C_{ee}$	&12.4	&4.6	&12.0	&5.4	&11.9	&4.4	&11.5	&5.2\\
$C_{le}$	&9.8	&4.0	&8.8	&4.2	&9.4	&3.9	&8.5	&4.0\\
$C_{W}$		&1.8	&4.5	&1.9	&4.5    &1.9	&4.4    &2.0	&4.5 \\
  \bottomrule
 \end{tabular}
\caption{Best fit values and corresponding $1\sigma$ confidence regions for $\Delta_{\rm SMEFT}=\{0\%, 1\%\}$ and for the two input parameter schemes considered in this work. These numbers have been obtained minimizing the $\chi^2$ with one parameter at a time (despite the non-minimal character of the SMEFT \cite{Jiang:2016czg}), and they have been multiplied by a factor 100.}\label{tab.fitresults_numeric_oneatatime_cll1}
\end{table}

\providecommand{\href}[2]{#2}\begingroup\raggedright\endgroup
\end{document}